\documentclass[12pt,fleqn, a4paper]{article}

\newcommand{\ra}{\rightarrow}
\newcommand{\tr}{\mbox{Tr}}
\newcommand{\bra}{\langle} \newcommand{\ket}{\rangle}
\newcommand{\be}{\begin{equation}}
\newcommand{\ee}{\end{equation}}
\newcommand{\bea}{\begin{eqnarray}}
\newcommand{\eea}{\end{eqnarray}}
\newcommand{\eps}{\epsilon}

\newcommand{\e}{\mbox{\scriptsize e}}
\newcommand{\ffi}{\varphi}

\newcommand{\ep}{\qquad {\vrule height 10pt width 8pt depth 0pt}}

\newcommand{\grintl}{[\kern-.18em [}
\newcommand{\grintr}{]\kern-.18em ]}

\newtheorem{lem}{Lemma}[section]
\newtheorem{prop}{Proposition}[section]
\newtheorem{thm}{Theorem}[section]
\newtheorem{cor}{Corollary}[section]
\usepackage{amsfonts}
\def\R{{\mathbb R}}
\def\hil{{\cal H}}
\def\ban{{\cal B}}
\def\0{{\mathbb O}}
\def\un{{\mathbb I}}

\def\Z{{\mathbb Z}}
\def\N{{\mathbb N}}
\def\C{{\mathbb C}}

\def\B{{\mathbb B}}

\begin{document}

\title{Weak Coupling and Continuous Limits for \\ 
Repeated Quantum Interactions
}

\author{St\'ephane Attal  \\ \\ Institut Girard Desargues\\ Universit\'e de Lyon 1\\
21 Av. Claude Bernard\\ 69622 Villeurbanne Cedex,\\ France
 \and 
Alain Joye \\ \\
Institut Fourier
\\ Universit\'e de Grenoble 1,\\ BP 74,
\\38402 St.-Martin d'H\`eres Cedex, \\ France}

\date{  }
\maketitle
\abstract{We consider a quantum system in contact with a heat bath
consisting in an infinite chain of identical sub-systems at thermal
equilibrium at inverse temperature $\beta$.
The time evolution is discrete and such that over each 
time step of duration $\tau$, the reference system is coupled to 
one new element of the chain only, by means of an interaction of 
strength $\lambda$. We consider three asymptotic regimes of the
parameters $\lambda$ and $\tau$ for which the effective 
evolution of observables on the small system becomes continuous 
over suitable macroscopic time scales $T$ and whose generator 
can be computed:
the weak coupling limit regime $\lambda\ra 0$, $\tau=1$, the regime 
$\tau\ra 0$, $\lambda^2\tau \ra 0$ and the critical case $\lambda^2\tau=1$, 
$\tau\ra 0$. The first two regimes are perturbative in nature and
the effective generators they determine is such that a non-trivial
invariant sub-algebra of observables naturally emerges.
The third asymptotic regime goes beyond the perturbative 
regime and provides an effective dynamics governed 
by a general Lindblad generator naturally constructed from the 
interaction Hamiltonian. Conversely, this result shows that
one can attach to any  Lindblad generator a repeated  quantum
interactions model whose asymptotic effective evolution is generated
by this Lindblad operator.
}

\setcounter{equation}{0}
\section{Introduction}

This paper is concerned with the study of the weak coupling limit, and variations thereof,
of open quantum systems consisting in a small quantum system defined by a Hamiltonian
$h_0$ on a Hilbert space $\hil_0$ coupled to a field or heat bath modelled by an 
infinite chain of identical independent $n+1$-level sub-systems on 
$\otimes_{\N^*}\hil$, with $n$ finite. The coupling between 
the distinguished system and the chain is provided by a discrete sequence of 
interactions of the small system with one individual sub-system of the chain, in the
following way:  if $\tau>0$ is a microscopic time scale, over a macroscopic time interval 
$]0,k\tau]$, $k\in\N^*$, the small system is coupled with elements 
$1,2, ..., k$ of the chain, in sequence,
for the same time $\tau$ and with the same interaction of strength $\lambda$. 
The interactions we consider are of the linear minimal coupling type 
$\sum_{j=0}^nV_j^*\otimes a_j+V_j\otimes a_j^*$, where the $a_j^*$'s and $a_j$'s are 
creation and annihilation operators relative to the levels of the sub-system 
and the $V_j$'s are arbitrary operators on $\hil_0$. Such models of repeated 
quantum interactions are used in physics, e.g. in quantum optics,
in the theory of quantum measurement or in decoherence. 
The lack of coupling, and thus of coherence, between the elements of the chain 
allows to expect that an effective continuous dissipative dynamics for pure states or 
observables on the small system of the form $e^{t\Gamma}$ should emerge when the 
number $k$ of discrete interactions goes to infinity and 
the coupling $\lambda$ with the chain elements is weak, in the familiar 
weak coupling regime. 
Recall that this corresponds to choosing  $t\in\R$ and considering $\N\ni k=t/\lambda^2$ 
so that the macroscopic time scale equals $T=\tau t/\lambda^2$. When $\tau$ is fixed 
and $\lambda\ra 0$, both $k$ and $T$ go to infinity as $1/\lambda^2$.
Moreover, in the setting adopted here, we have another parameter at hand
which is the microscopic interaction time $\tau$ of  the small system with 
each individual element of the chain. 
It allows us to explore different asymptotic regimes, as $\tau$ goes to 
zero as well, which characterizes the continuous limit, over suitable
macroscopic time scales $T$.

One goal of this paper is to establish the existence of effective 
continuous Markovian dynamics in weak and/or continuous limits
defining three asymptotic regimes. 
We consider successively the effective Schr\"odinger evolution
on the small system at zero temperature and, when the chain is at equilibrium at
zero or positive temperature, the Heisenberg evolution of observables on the
small system. While the existence of an effective dynamics obtained by a weak 
limit procedure ($\tau=1$) is proven for a large class of time-independent 
Hamiltonian systems, as well as in certain  time-dependent situations, see e.g. 
\cite{d0}, \cite{d}, \cite{ds}, \cite{ls}, \cite{dj},  this question is not 
addressed in the literature for the case under study. 
Note also that a Hamiltonian formulation of our system 
necessarily involves a piecewise constant time-dependent generator. 
The analysis relies on the following property of the model, which is inherent to its
definition. The effective dynamics on the small system of pure states or of observables
from time $0$ to time $k\tau$ is shown to be given by the $k^{\mbox{ th }}$ power of a 
linear operator, which depends on the parameters $\lambda$ and $\tau$. This
expresses the Markov property in a discrete setting.

The first part of the paper is devoted to the usual weak limit regime $\lambda\ra 0$, 
$\tau$ fixed and $T=t\tau/\lambda^2\ra\infty$, $0<t$ finite. 
We show the existence an effective dynamics driven by a $\tau$ dependent generator
which we determine. This first result is obtained by adapting the arguments
developed in the study of the weak coupling regime for stationary Hamiltonians
to our discrete quantum dynamics framework. 
The method is then extended 
to accommodate the whole range $\tau\ra 0$, $\lambda^2\tau\ra 0$
over macroscopic time scales $T=t/(\tau\lambda^2)\ra \infty$, which
defines our second regime. This
gives rise to an effective dynamics driven by a $\tau$ independent generator
we compute as well and to which we come back below.
The analysis of these first two regimes is 
strongly related to regular perturbation theory in the parameter $\lambda^2\tau$
and we refer to these regimes as perturbative regimes. Technically, the 
study of the second regime relies on an asymptotic analysis in the two parameters 
$\lambda$ and  $\tau$ of the discrete evolution of our system.
The divergence of the macroscopic time scale imposes, as usual, some renormalization of
the dynamics by the restriction of the uncoupled dynamics. 
Finally, note that in the second regime,
the interaction strength  $\lambda$ is not required to go to zero and can even diverge.
The common feature of the generators of the dynamics of observables obtained in 
these two regimes is that they commute with the generator $i[h_0,\cdot ]$ of the uncoupled 
unitary evolution restricted to $\hil_0$. In other words, 
the corresponding effective dynamics admits the commutant of $h_0$ as a non trivial 
invariant sub-algebra of observables. This property is well known in the weak 
coupling regime for time-independent Hamiltonians, \cite{d}, \cite{ls}, \cite{dj}.

Our primary motivation actually comes from the recent paper \cite{ap} where
such repeated interactions models are shown to converge in some subtle limiting 
procedure to open quantum systems with a heat bath consisting in continuous 
fields of quantum noises, at zero temperature. 
These limiting systems give rise in a natural 
and spontaneous way to effective dynamics on the Hilbert space $\hil_0$ of 
the small system governed by quantum Langevin equations. 
The above mentioned limit involves at the same time the time scale $\tau$, 
the strength of interaction $\lambda$ as well as a notion of spacing between 
the sub-systems forming the chain in an intricate way. While reminiscent of 
weak coupling  methods in spirit, the 
limiting procedure of \cite{ap} is nevertheless distinct from the weak
coupling limit. Indeed, while $\tau\ra 0$, the product $\lambda^2\tau$ is kept
constant in \cite{ap},  which leads us beyond the perturbative regime.
Another goal of the present work is to consider as our third regime the 
critical scaling  
$\lambda^2\tau=1$ in our repeated quantum interactions model and to derive 
an effective evolution of observables for a chain at inverse temperature 
$\beta$. The relevant macroscopic time scale in this regime is $T=t/(\lambda^2\tau)=t$, 
which is finite. 

With this scaling, we show that an effective Heisenberg 
dynamics for observables  on $\hil_0$  emerges at any temperature.
It is generated by a general Lindblad operator 
whose dissipative part is explicitly constructed in terms of the
$V_j$'s defining the coupling in the Hamiltonian, whereas its conservative 
part is simply $i[h_0,\cdot]$. At zero temperature, we
recover  the effective Heisenberg dynamics of observables on $\hil_0$ 
of \cite{ap} obtained by means of quantum noises. At positive temperature, 
our generator coincides with a construction proposed in
\cite{lm} for certain models using an a priori modelization of the heat 
bath by some thermal quantum noises, generalizing those used at zero temperature. 
For any temperature, the effective dynamics is distinct from that obtained in the 
previous two perturbative regimes.
In particular, the generator obtained does not commute with $i[h_0,\cdot]$ anymore, 
the generator of the uncoupled evolution restricted to $\hil_0$. Hence, there is 
no obvious sub-algebra of observables  left invariant  by the effective dynamics
of observables. The analysis of this critical case makes 
use of Chernoff's Theorem, rather than perturbative methods.

Let us compare the generator of the effective dynamics of observables obtained in the regime 
$\tau\ra 0$, $\tau\lambda^2\ra 0$, and the general Lindblad operator obtained as 
$\tau\ra 0$ with $\tau\lambda^2=1$. In the  former case, the generator is 
obtained from the dissipative part of the Lindblad operator of the latter case 
by retaining its diagonal terms only with respect to the spectral 
decomposition of the uncoupled evolution restricted to $\hil_0$. Or, in an equivalent
way, by performing a time average of the Lindblad operator with respect to the 
uncoupled evolution restricted to $\hil_0$.
This defines the so-called $\#$ operation that makes the 
commutant of $h_0$ invariant under the effective dynamics in the regime 
$\tau\ra 0$, $\tau\lambda^2\ra 0$ (and in the weak coupling regime as well). 
Our results show that the $\#$ operation is present as long 
as $\lambda^2\tau\ra 0$, whereas it disappears in the critical regime 
$\tau\lambda^2$. In other words, in the regime $\tau\ra 0$, $\tau\lambda^2\ra 0$, 
a non-trivial distinguished invariant sub-algebra of observables exists, 
whereas in the critical case  $\tau\ra 0$, $\tau\lambda^2=1$, there is a priori 
no sub-algebra left invariant by the effective dynamics, since its generator takes 
the form of a generic Lindblad operator.

We finally note here that from a practical point of view, the modelization of 
the dynamics of observables (or states) of a small system in contact with a 
reservoir at a certain temperature often starts with a choice of a certain 
Lindblad generator suited to the physical phenomena to be discussed. 
Our analysis allows to assign to any Lindblad generator 
a simple model of repeated quantum interactions, with explicit couplings constructed
from the Lindblad generator, whose effective dynamics in the limit 
$\tau\ra 0$, $\lambda=1/\sqrt{\tau}$, is generated by the chosen Lindblad operator.

\vskip.5cm 

The paper is organized as follows. The general setup and 
definition of the model are provided in the next section. 
Section 3 is devoted to the analysis of the weak limit of the model 
at zero temperature, in the Schr\"odinger picture. Our main results 
in this setup are expressed as Corollary \ref{c2} for the weak coupling 
regime and Corollary \ref{3.5} for the regime $\lambda^2\tau\ra 0$, 
$\tau\ra 0$. This section also contains the technical basis underlying 
our perturbative analysese in both the Schr\"odinger and Heisenberg pictures.
The main technical result, of independent interest, is actually valid in 
a Banach space framework and  
is stated as Theorem \ref{theo31}. The positive temperature case,
in the Heisenberg picture is dealt with in Section 4. The generators of the
effective dynamics of observables in the two perturbative regimes are given
in Theorem \ref{mtb} and Corollary \ref{tobewritten}. The analysis of the
critical regime $\lambda^2\tau=1$ is presented in Section 5, for both
the Schr\"odinger and Heisenberg pictures. Section 6 is devoted
to a thorough analysis of the first non-trivial case where both the small system and
the elements of the chain consist in two-level systems.

\setcounter{equation}{0}
\section{A Repeated Interaction Model}

Consider the following setup to start with. Our small system, described by the Hilbert
space $\hil_0$ of dimension $d+1>1$ and a self-adjoint Hamiltonian $h_0$,
interacts with an infinite chain of identical finite dimensional sub-systems modelling 
a field or heat bath, by means of a time dependent Hamiltonian. 
The total Hilbert space is $\hil_0 \otimes \hil $, where $\hil=\otimes_{j\geq 1}\C^{n+1}$,
$n\geq1$. We will call the $j$th Hilbert space $\C^{n+1}_j\equiv \C^{n+1}$, the Hilbert 
space at site $j$,
$j=1,2, \cdots$ and, following the usage when $n=1$, we will call the subsystem at site $j$ 
the spin at site $j$. 
We adopt the following convenient notations used in \cite{ap}. The vacuum $\Omega\in \hil$ 
is defined as the infinite tensor product of the vacuum vector 
$\omega= \pmatrix{0 &\cdots& 0 & 1}^T$ in $\C^{n+1}$,
\be
\Omega=\omega \otimes \omega \otimes\omega \otimes  \cdots \ \in \  
\C^{n+1}\otimes \C^{n+1}\otimes \C^{n+1}\otimes \cdots .
\ee
Denoting the $i$th excited vectors $x_i=\pmatrix{0 & \cdots & 0&1 & 0&\cdots& 0}^T$, where the
$1$ sits at the $i$th line, starting from the bottom, $i=1,2, \cdots , d$,
the corresponding excited state at site $j\geq 1$ is given by 
\be
x_i(j)=\omega \otimes \cdots  \otimes\omega \otimes x_i
\otimes\omega \otimes  \cdots ,
\ee
where $x_i$ sits at site $j\geq 1$.
More generally, given a finite set 
\be\label{defs}
S=\{ (k_1, i_1), (k_2,i_2), \cdots, (k_m, i_m) \} \subset 
(\N^*\times \{1,2, \cdots , d\})^m \ \ \mbox{ with all } k_j\mbox{'s distinct},
\ee 
we define $X_{S}$ as the vector 
given by an infinite tensor product as above, with $i_j$th excited vectors $x_{i_j}(k_j)$
at all sites $k_j\geq 1$, $j=1,\cdots, m$,  and ground state vectors $\omega$ everywhere else.
This construction together with the vacuum $\Omega\equiv X_\emptyset$ yield an orthonormal basis 
 of $\hil$, when $S$ runs over all finite sets of the type above.

Let us introduce creation and annihilation operators associated with the vectors $x_i(j)$.
Let $a_i$ and $a_i^*$, $i=1,2,\cdots, n$, denote the operators corresponding to 
$\{\omega, x_1, \cdots , x_n\}$ in $\C^{n+1}$, i.e. such that
\bea
& & a_i x_i=\omega, \ \ a_i \omega =a_i x_j =0, \ \mbox{ if } \ j\neq i,\nonumber\\
& & a_i^* \omega =x_i, \ \ a_i^* x_j =0 \ \mbox{ for any } \ j=1,2,\cdots, n.
\eea
Note that these operators do not coincide with the familiar creation and 
annihiliation,
however, for $i$ fixed, they satisfy the anti-commutation rules when restricted to the
two dimensional subspace $<\omega, x_i>$ and are zero on the orthogonal complement 
of this subspace. Then, for $j\geq 1$, the operators $a_i(j)$ and $a_i(j)^*$ on 
$\hil$ are defined as acting as $a_i$ and $a_i^*$ on the $j$th copy of 
$\C^{n+1}$ at site $j$, and as the identity everywhere else. Therefore, 
when acting on different copies of $\C^{n+1}$, these operators commute.
In keeping with the notations for the reservoir, we introduce a basis of 
eigenvectors of $h_0$  for $\hil_0$ of the form
\be
\{\omega, x_1, x_2, \cdots, x_d\}, \ \ \mbox{ where } \ d=\dim(\hil_0)-1.
\ee
Note that  $d\neq n$ in general, but we shall nevertheless use sometimes
the notation $\omega(0)$ and $\{x_i(0)\}_{i=1,2, \cdots, d}$ to denote these vectors. 
No confusion should arise with vectors of $\hil$ above, since we labelled the sites
of the spins by positive integers. In some cases, $\hil_0$ will be an infinite 
dimensional separable Hilbert space, which corresponds formally to $d=\infty$.

\vskip.3cm

Our formal time dependent Hamiltonian $H(t,\lambda)$ on $\hil_0\otimes\hil$
has the form
\be\label{hamil}
H(t,\lambda)=H_0+H_F+\lambda H_I(t), 
\ee 
where
\be\label{hamfield}
H_0=h_0\otimes \un , \ \ \ H_F=\sum_{j=1}^\infty \sum_{i=1}^n \un 
\otimes \delta_i a_i(j)^* a_i(j),
\ \mbox{ with } \  \delta_i\in\R, 
\ee
and, for $t\in [\tau (k-1) ,\tau k[$, 
\be\label{defint}
H_I(t)=\sum_{i=1}^n V^*_i\otimes a_i(k)+V_i\otimes a_i(k)^*\equiv I(k), 
\ee
where the $V_i$'s and $h_0$ are bounded operators on $\hil_0$, in case $\hil_0$ is a 
separable infinite dimensional Hilbert space.
These operators describe the 
interaction between the small system with the different levels of energy $\delta_i$ of the spin
at site $k$, during the time interval $]\tau (k-1) ,\tau k]$ of length $\tau$. The form of
$H_F$ makes it an unbounded operator, but, as we will see in the sequel, we will only make use
of the unitary evolution it generates and, moreover, it will always be sufficient to work with
subspaces containing finitely many excited states only. 

In order to make the notations more compact, we introduce vectors with operator
valued entries that allow to get rid of the indices $i=1,\cdots, n$. Let
\bea
&&a(j)^{\sharp}=\pmatrix{a_1(j)^{\sharp} & a_2(j)^{\sharp} & \cdots & a_n(j)^{\sharp}}^T \\
&&V^\sharp=\pmatrix{V_1^\sharp & V_2^{\sharp} & \cdots & V_n^{\sharp}}
\eea
where $^{\sharp}$ denotes either nothing or $*$. Then, using the rules of matrix composition, 
we can write
\be
V^{\sharp_1}\otimes a(j)^{\sharp_2}=\sum_{i=1}^n V^{\sharp_1}_i\otimes a_i^{\sharp_2}(j),
\ee
so that we can rewrite the interaction Hamiltonian for $t\in ]\tau (k-1) ,\tau k]$ as 
\be
I(k)=V^*\otimes a(k)+V\otimes a(k)^*.
\ee
Similarly, with
\bea
&&a(j)^{\sharp}a(j)=\pmatrix{a_1(j)^{\sharp}a_1(j) & a_2(j)^{\sharp}a_2(j) & \cdots & 
a_n(j)^{\sharp}a_n(j)}^T \\
&&\delta = \pmatrix{\delta_1  & \delta_2  & \cdots & \delta_n },
\eea
we can write
\be
H_F=\un \otimes \sum_{j\geq 1}\delta  a(j)^*a(j).
\ee

\vskip.3cm

We will denote the corresponding evolution operator
between the time $\tau (k-1)$ and $\tau k$ by $U_k$, so that
\be\label{brick}
U_k=e^{-i\tau (H_0+ H_F+ \lambda I(k))},
\ee
and the evolution from $0$ to $\tau n$ is given by 
\be\label{evol}
U(n,0)=U_nU_{n-1} \cdots U_k\cdots U_1.
\ee
Although not explicited in the notation, the operator
$ U(n,0)$ depends on $\lambda$ and $\tau$. \\

\vskip.3cm

We will first be interested in the weak coupling limit of this evolution operator 
characterized by the familiar  scaling 
\be\label{scale}
n=t/\lambda^2 \  , \ \ \ \lambda\ra 0 \ \ \ \mbox{ and } \ \ \ \tau \ \ \mbox{fixed.} 
\ee
Hence, the macroscopic time scale $T$ is given by
\be\label{macro}
T=\tau n = \tau t/\lambda^2 \ra \infty.
\ee

Note, however, that in contrast with the usual set up, we have here a non-smooth time 
dependent Hamiltonian $H(t,\lambda)$.\\

\noindent
{\bf Remarks:}\\
i) In order not to bury the main points of our analysis under technical subtleties,
we have chosen to work in a simple framework where all relevant operators are
bounded or matrix valued. Nevertheless, some of our results below hold if we consider
our heat bath to live in a tensor product of infinite dimensional separable Hilbert 
spaces and make further assumptions so that the field Hamiltonian and interaction
are bounded. \\
ii) In some cases we shall allow $\hil_0$ to be a separable Hilbert space. 
This will be explicitly stated in the hypotheses. Otherwise, we will 
work on the model  defined above, under the general assumption \\

{\bf H0: } The Hamiltonian is defined on the Hilbert space $\hil_0\otimes \hil$,
where $\hil_0=\C^{d+1}$,  $\hil=\otimes_{j\geq 1}\C^{n+1}$, for $d,n$ finite, and is
given by (\ref{hamil}), (\ref{hamfield}), (\ref{defint}). The evolution it generates
is given by (\ref{evol}) and (\ref{brick}).

\section{Weak limit of the Schr\"odinger representation at zero temperature}
\setcounter{equation}{0}

As a warm up, and in order to derive some preliminary estimates, we
prove here the existence of the weak limit for our model at zero 
temperature in the Schr\"odinger picture, and compute this limit. 
We first prove a key lemma that reduces 
the computation of the projected part of the evolution $U(n,0)$ (\ref{evol}) to the
$n^{\mbox{ th }}$ power of a single matrix. Then we perform a general analysis of 
large powers of operators based on perturbative expansions which appear in
the computations of weak limits. These technical results are expressed in 
Proposition \ref{1p} and Theorem \ref{theo31} under different sets of hypotheses. Their 
applications to our model are given in Corollaries \ref{c2} and  \ref{3.5}.
\\

\subsection{Markov Properties}

Let $P$ be the projection from $\hil_0\otimes \hil$ to the subspace $\hil_0\otimes \C \Omega$
defined by
\be
P=\un \otimes |\Omega\ket\bra\Omega|.
\ee
The object of interest to us in this Section will  thus be the limit
\be
 \lim_{\lambda\ra 0}P(U(t/\lambda^2,0))P,
\ee
as an operator from $P \ \hil_0\otimes \hil$ to $P \ \hil_0\otimes \hil$, identified with 
$\hil_0$, the Hilbert space of the small system.

Note that 
\be
U_j=e^{-i\tau \hat{H}_j}e^{-i\tau  \tilde{H}_j},
\ee
with
\bea\label{H(0)}
& &\widetilde{H}_j=h_0\otimes \un  + \un \otimes \delta a(j)^* a(j)
+\lambda (V^* \otimes a(j)+V\otimes a(j)^*) \nonumber\\
& &\widehat{H}_j =\un \otimes \sum_{k\neq j} \delta  a(k)^* a(k),
\eea
two operators that commute.

We observe the following property of products of operators $U_k$, which shows the
Markovian nature of the reduced evolution.
\begin{lem}\label{red}
Let us write the restriction of $U_j$ to $\hil_0\otimes \C^{n+1}_j$ 
as a block matrix with respect to the ordered basis of 
$\hil_0\otimes \C^{n+1}_j$
\be\label{ordb}
\matrix{\{\omega \otimes \omega, x_1 \otimes \omega, \cdots, 
 x_d \otimes \omega, \cr \phantom{\{x }
\omega \otimes x_1, x_1\otimes x_1, \cdots  x_d \otimes 
x_1,\cr \vdots \cr \phantom{\{xx }
\omega \otimes x_n, x_1\otimes x_n, \cdots  x_d \otimes 
x_n\}}
\ee
as 
\be\label{decomp}
U_j|_{\hil_0\otimes \C^{n+1}_j}=\pmatrix{A & B \cr C & D},
\ee
where $A$ is a $(d+1)\times (d+1)$ matrix, $B$ is $(d+1)\times n(d+1)$, $C$ is
$(d+1)n\times (d+1)$ and $D$ is $n(d+1) \times n(d+1)$. Then, for any $m\geq 0$,
\be
P U(m,0)P= A^m\otimes |\Omega\ket\bra \Omega |\simeq A^m.
\ee
\end{lem}
{\bf Proof:} Follows from the fact that
\be
U_j (\un \otimes |\Omega\ket\bra\Omega |)=
e^{-i\tau  \widetilde{H}_j}(\un \otimes |\Omega\ket\bra\Omega |)
\ee
where, if $\hil_0\ni v=v_0\omega(0) +\sum_{i=1}^dv_i x_i(0)\simeq \vec v$,
\be
e^{-i\tau  \widetilde{H}_j} v\otimes\Omega =
A\vec v\otimes \Omega+\sum_{i=1}^{n+1}(C\vec v)_i\otimes x_i(j),
\ee
where $(\vec w)_i$ denotes the $i$'th component of the vector $\vec w$. 
Hence, due to the fact that different $U_j$'s act on different $\C_j^{n+1}$'s,
\be
U_mU_{m-1}\cdots U_1 \, v\otimes\Omega= A^m\vec v\otimes \Omega + 
\sum_{i=1}^{m(d+1)}\vec w_i\otimes X_{S_i},
\ee
where $\vec w_i$ are some vector in $\C^{d+1}$ and the excited sets $S_i$ are
never empty.
Therefore their contribution vanishes in the computation 
\be
(\un \otimes |\Omega\ket\bra\Omega|) U_mU_{m-1}\cdots U_1 \,  
v\otimes\Omega=
 A^m\vec v\otimes |\Omega\ket\bra\Omega|.
\ee
\hfill \ep\\

As is easy to check along the same lines, in case $\hil_0$ is infinite dimensional, 
we can generalize the above Lemma as follows.
\begin{lem}\label{upgrade}
Let $\hil_0$ be a separable Hilbert space and $h_0$, $V_j$, $j=1,\cdots, n$ be bounded
on $\hil_0$.
We set \be
P_j=|x_j\ket\bra x_j | : \C^{n+1}\mapsto \C^{n+1}, \, \, j=1,\cdots, n, \,\,  
P_0 =|\omega \ket\bra \omega |\, \, \mbox{ and } \,\,Q_0=\un -P_0,
\ee
so that 
\be
\hil_0\otimes \C^{n+1}=(\hil_0\otimes P_0\C^{n+1})\, \oplus \, (\hil_0\otimes Q_0\C^{n+1})
\simeq (\hil_0\otimes \C )\,  \oplus \, (\hil_0\otimes \C^{n}).
\ee
We can decompose
\be\label{decompup}
U_j|_{\hil_0\otimes \C^{n+1}_j}=\pmatrix{A & B \cr C & D},
\ee
where $A: \hil_0 \mapsto \hil_0$, $B:  \hil_0\otimes\C^n \mapsto \hil_0$,
 $C:  \hil_0 \mapsto \hil_0\otimes\C^n$
and $D:  \hil_0\otimes\C^n \mapsto \hil_0\otimes\C^n$. Then, for any $m\geq 1$,
\be
P U(m,0)P= A^m\otimes |\Omega\ket\bra \Omega |\simeq A^m.
\ee
\end{lem}

The above Lemmas thus lead us to consider a reduced problem on $\hil_0$.
We need to compute the matrix $A$ in the decomposition (\ref{decomp})
of $e^{-i\tau(h_0+\delta a^*a+\lambda (V^* a +V a^*))}$, where we
dropped the indices $j$, the $\un$ and the $\otimes$ symbol in the notation.
Recall however that a summation over the excited states of $H_F$ is
implicit in the notation. \\

\vskip.3cm

\subsection{Preliminary Estimates}
In order to apply perturbation theory as $\lambda\ra 0$ and, later on,
in other regimes involving $\tau\ra 0$ as well, we derive below estimates 
to be used throughout the paper.

We  rewrite
the generator as
\be
H(\lambda)=H(0)+\lambda W, \ \ \ \mbox{with} \ \ \ H(0)=h_0+\delta a^*a 
\ \ \mbox{and } \ \ W=V^* a +V a^*.
\ee
With a slight abuse of notations, the projector $P$ takes the form 
\be
P=\un- a^*a .
\ee
We can slightly generalize the setup and work under the following
hypothesis:\\

\noindent
{\bf H1:}\\
Let $P$ be a projector on a Banach space $\ban$ and $H(\lambda)$ be 
an operator in of the form
\be
H(\lambda)=H(0)+\lambda W,
\ee
where $H(0)$ and $W$ are bounded and $0\leq\lambda\leq\lambda_0$ for
some $\lambda_0>0$.  Further assume that
\be
[P,H(0)]=0\ \ \ \mbox{and }\ \  W=PWQ+QWP \ \ \mbox{where }\ \ \ Q=\un -P.
\ee
\vspace{.2cm}

\noindent
We consider 
\be\label{lop}
U_\tau(\lambda)=e^{-i\tau H(\lambda)}. 
\ee
For later purposes, we also take care of the dependence in $\tau$ of the 
error terms. As this parameter will eventually tend to zero in some
applications to come below, we consider the error terms as both $\lambda$ and 
$\tau$ tend to zero, independently of each other. We have a first easy perturbative result
\begin{lem}\label{easypert}
Let {\bf H1} be true. Then, as $\lambda$ and $\tau$ go to zero, 
\bea \label{ff}
&&e^{-i\tau(H(0)+\lambda W)}=e^{-i\tau H(0)}+\lambda F(\tau)+\lambda^2 G(\tau)
+O(\lambda^3\tau^3)\\
\label{pup}
&&Pe^{-i\tau(H(0)+\lambda W)}P=Pe^{-i\tau H(0)}P+\lambda^2 PG(\tau)P+PO(\lambda^4\tau^4)P,
\eea
where 
\bea\label{deff}
F(\tau)&=&\sum_{n\geq 1}\frac{(-i\tau)^n}{n!}
\sum_{m_j\in \N \atop m_1+m_2=n-1}H(0)^{m_1}WH(0)^{m_2}  \nonumber\\
&=&-i e^{-i\tau H(0)}\int_0^\tau ds_1 e^{i s_1 H(0)}We^{-i s_1 H(0)}\\
\label{defg}
G(\tau)&=&\sum_{n\geq 2}\frac{(-i\tau)^n}{n!}\sum_{m_j\in \N \atop m_1+m_2+m_3=n-2}
H(0)^{m_1}WH(0)^{m_2}WH(0)^{m_3}\nonumber\\
&=&- e^{-i\tau H(0)}\int_0^\tau ds_1 \int_0^{s_1} ds_2  e^{i s_1 H(0)}W e^{-i (s_1-s_2) H(0)}
W e^{-i s_2H(0)}.
\eea
Moreover
\bea
&&\frac{d }{d\tau } G(\tau)=-iH(0)G(\tau)-iWF(\tau), \,\,\, G(0)=0\\
&&F(-\tau)=-e^{i\tau H(0)}F(\tau)e^{i\tau H(0)}\\
&&G(-\tau)=-e^{i\tau H(0)}G(\tau)e^{i\tau H(0)}+
e^{i\tau H(0)}F(\tau)e^{i\tau H(0)}F(\tau)e^{i\tau H(0)}.
\eea
\end{lem}
{\bf Remark:} Formula (\ref{ff}) is true without assuming that
$W$ is off-diagonal with respect to $P$ and $Q$. 

\vspace{.3cm}

\noindent
{\bf Proof:} First note that $U_\tau(\lambda)=e^{-i\tau H(\lambda)}$ is analytic
in both variables $\lambda$ and $\tau$ in $\C^2$. Then, we 
compute the exponential of $-i\tau$ times $H(\lambda)$ as a convergent series.
Consider terms of the form
\bea\label{exp}
(H(0)+\lambda W)^n&=&H(0)^n+\lambda \sum_{k=0}^{n-1}H(0)^kWH(0)^{n-1-k}\nonumber\\
&+&\lambda^2
\sum_{m_j\in \N \atop m_1+m_2+m_3=n-2}H(0)^{m_1}WH(0)^{m_2}WH(0)^{m_3}+O(\lambda^3 C^n).
\eea
The error term in $C^n$ comes from the boundedness of the operators involved. 
Multiplication by $(-i\tau)^n/n!$ and summation over $n\geq 0$ yields the first 
result with our definition of $F(\tau)$ and $G(\tau)$. The second result follows
from taking into account that $PWP=QWQ=0$, hence only the terms with an even number 
of $W$'s survive and we get
\bea
&&P\frac{(H(0)+\lambda W)^n}{n!}P=\\\nonumber
&&P\left( \frac{H(0)^n}{n!}+\lambda^2
\sum_{m_j\in \N \atop m_1+m_2+m_3=n-2}\frac{H(0)^{m_1}WH(0)^{m_2}WH(0)^{m_3}}{n!}
+O(\lambda^4C^n/n!)\right)P.
\eea
The overall error in $\tau^4\lambda^4$ comes from the fact that it takes at least 
four terms in (\ref{exp}) to get a contribution of order $\lambda^4$.
The computation above was conducted to order $\lambda^2$ because of
the scaling (\ref{scale}). 
The order $\lambda$ term $F(\tau)$
doesn't contribute, being off diagonal with respect to $P$.
\\

An alternative derivation of a perturbation series of $e^{-i\tau(H(0)+\lambda
W)}$ in $\lambda$ yields the other expressions for $F(\tau)$ and $G(\tau)$.
It is obtained via Dyson series in the familiar interaction 
picture. We have the identity 
\be
 i\frac{d}{d\tau}e^{-i\tau(H(0)+\lambda W)}=(H(0)+\lambda W)e^{-i\tau(H(0)+\lambda W)},\,\,\,
\left.e^{-i\tau(H(0)+\lambda W)}\right|_{\tau=0}=\un.
\ee
Introducing 
\be
\Theta(\lambda, \tau)=e^{i\tau H(0)}e^{-i\tau(H(0)+\lambda W)},
\ee
this operator satisfies
\be
i\frac{d}{d\tau}\Theta(\lambda, \tau)=\lambda e^{i\tau H(0)}We^{-i\tau H(0)}
\Theta(\lambda, \tau), \,\,\,\left.\Theta(\lambda, \tau)\right|_{\tau=0}=\un.
\ee
Hence we have the convergent expansion
\bea\label{cv}
\Theta(\lambda, \tau)&=&\sum_{n=0}^{\infty} (-i\lambda)^n\int_0^\tau ds_1\int_0^{s_1} 
ds_2 \cdots \int_0^{s_{n-1}}ds_n e^{i s_1 H(0)}W\times \nonumber\\
& & \times e^{-i (s_1-s_2) H(0)}W e^{-i (s_2-s_3) H(0)}
\cdots e^{-i s_{n-1}-s_n) H(0)}We^{-i s_n H(0)}.
\eea
Therefore, focusing on the terms of order $\lambda$ and $\lambda^2$, we get the 
alternative expressions for $F(\tau)$ and $G(\tau)$.

The differential equation yielding $G(\tau)$ as a function of
$F(\tau)$ follows from explicit computations on the  expressions above, as the
identities for $\tau\mapsto -\tau$.\ep
 
Let us give some more properties of the expansion of $U_\tau(\lambda)$ for
$\lambda>0$ small, $\tau>0$ in the Hilbert space context that will be used later on.
\begin{cor}\label{pertuu} Assume $\ban$ is a Hilbert space, $H(0)$, $W$ and $P$
are self-adjoint and $\lambda, \tau $ are real.
As $\lambda\ra 0$, the operator
$U_\tau(\lambda)=e^{-i\tau H(\lambda)}$ satisfies
\bea
&&U_\tau(\lambda)=e^{-i\tau H(0)}+\lambda F(\tau)+\lambda^2 G(\tau)+O(\lambda^3\tau^3)\\
&&U_\tau(\lambda)^{-1}=U_\tau(\lambda)^*=U_{-\tau}(\lambda)\nonumber\\
&&\quad \quad \quad \quad =e^{i\tau H(0)}+\lambda F(-\tau)+\lambda^2 G(-\tau)+O(\lambda^3\tau^3)
\eea
with the identities for all $\tau\in\R$
\bea\label{ftau}
&&F(-\tau)=F^*(\tau)\\
\label{gtau}
&&G(-\tau)=G^*(\tau).
\eea
\end{cor}
{\bf Proof:} Follows from the fact that $H(\lambda)$ is self-adjoint. 
\hfill \ep\\

\subsection{Weak Limit Results}

The technical basis underlying all our weak limit results 
is contained in the next two Lemmas and the Proposition
following them. They are stated in a general framework that will
suit both our analysese of the Schr\"odinger and Heisenberg representations. 
This is why we use independ notations.

\begin{lem}\label{step1}
Let $V(x)$, $x\in [0,x_0)$, and $R$  be bounded linear 
operators on a Banach  space $\ban$ such that,  in the operator norm, 
$V(x)=V(0)+ x R + O(x^2)$, and $V(0)$ is an isometry which admits 
the following spectral decomposition  
\be\label{ldb}
V(0)=\sum_{j=0}^{r}e^{-iE_j}P_j \ \ \ \mbox{ where } \ \  r<\infty ,\ \
E_j\in\R, \ \  \{e^{-iE_j}\}_{j=0,\cdots , r} \ \ \mbox{distinct.} 
\ee
Let $h=\sum_{j=0}^{r}E_jP_j$ so that $V(0)=e^{-ih}$ and
\be\label{defj}
J=\sum_{j,k=0}^{r}\alpha_{jk}P_jRP_k \ \ \ \mbox{ where }
\ \ \ \alpha_{jk}=\left\{\matrix{\frac{E_j-E_k}{e^{-iE_j}-e^{-iE_k}} & 
\mbox{ if } \ j\neq k 
\cr ie^{iE_j} &\mbox{ if }\ j= k. }\right. 
\ee
Then, for any $0\leq t\leq t_0$, where $t_0$ finite, and  $t/x\in \N$, 
\be
\|V(x)^\frac{t}{x} - {e^{-i(h+xJ)}}^\frac{t}{x} \|=O(x) , \ \ \mbox{ as }\ \ x\ra 0, 
\ \ \mbox{ s.t. } \ \ t/x\in \N. 
\ee
\end{lem}

\noindent
{\bf Remarks:} \\
i) Expressing the projectors $P_j$ by Von Neumann's ergodic theorem as
\be
P_j=\lim_{N\ra\infty}\frac{1}{N}\sum_{n=0}^{N-1}(e^{iE_j}V(0))^n
\ee
shows that they are of norm one.\\
ii) The operator $J=J(R,h)$ is defined as the solution to the equation (\ref{lex}).
This equation is a particular case of  $i\int_0^1\ e^{ish}Xe^{-ish} \ ds = Y$
which is solved in a similar fashion.\\

\noindent
{\bf Proof:} With $m=t/x\in \N$,
\be
V(x)^m - {e^{-i(h+xJ)}}^m =\sum_{k=0}^{m-1}V(x)^k(V(x)-e^{-i(h+xJ)})
{e^{-i(h+xtJ)}}^{m-1-k}
\ee
where, by hypothesis and Lemma \ref{easypert}
\be\label{lex}
V(x)-e^{-i(h+xJ)}=x\left(R+ie^{-ih}\int_0^1 \ e^{ihs}Je^{-ihs}\ ds\right)
+O(x^2).
\ee
Moreover, note also
\be
\|V(x)\|= 1+O(x), \ \ \ \|e^{-i(h+xJ)}\|= 1+O(x).
\ee
Our definition (\ref{defj}) of $J$ is designed to make the term of
order $x$ in (\ref{lex}) vanish. Therefore, there exists  positive constants 
$c_0, c_1$ such that we can estimate for any $0\leq t\leq t_0 <\infty$
\bea
&&\| V(x)^m - {e^{-i(h+xJ)}}^m \| \leq c x^2 \sum_{k=0}^{m-1}\|V(x)\|^k
\|e^{-i(h+xJ)}\|^{m-1-k}\\
& &\quad \quad \quad \quad  \quad \quad \quad \leq  c_0 x^2 m ( 1+c_0x)^m\leq c_0txe^{\frac{t}{x}\ln (1+c_0x)}
 \leq  x c_0 t_0 e^{c_1 t_0}=O(x). \ep\nonumber
\eea

It will be necessary to control the dependence of such
estimates on a parameter $\tau\ra 0$ later on. This will cause no 
serious difficulty, since all steps are explicited in the argument.
To achieve sufficient control in $\tau$, we need to revisit
the proof of a well known lemma, which holds under weaker hypothesese than
ours, see Davies \cite{d}.
\begin{lem}\label{step2} Let $e^{-ih}=\sum_{j=0}^{r}e^{-iE_j}P_j$ 
be the isometry (\ref{ldb}) on 
the Banach space $\ban$ and let $K$ be a bounded operator on $\ban$. 
There exists a constant $c$ depending on $r$ and $t_0$ only, such 
that for any $t\in [0,t_0]$, $t_0$ finite, 
\be
\|e^{ith/x}e^{-i\frac{t}{x}(h+xK)}-e^{-itK^\#}\|\leq 
c\frac{x \|K\|(1+\|K\|)e^{2\|K\|t_0}}{\inf_{j\neq k}|E_j-E_k|}, 
\ \ \mbox{ as }  
\ \ \ x\ra 0,
\ee
 where $K^\#=\sum_{j=0}^r P_jKP_j=
\lim_{T\ra\infty}\frac{1}{T}\int_0^Te^{i s h}Ke^{-i s h}\, ds$.
\end{lem}
{\bf Remark:}  The expression of $K^\#$ as a Cesaro mean
is a classical computation which shows that $\|K^\#\|\leq \|K\|$.

\vspace{.3cm}

\noindent
{\bf Proof:} We follow \cite{d}. Let $f\in \ban$ and 
\be
f_x(t)=e^{ith/x}e^{-i\frac{t}{x}(h+xK)}f, \ \ \ f(t)=e^{-itK^\#}f.
\ee
By the fundamental Theorem of calculus, we can write
\bea
& &i(f_x(t)-f(t))=\int_0^t\ \left(e^{ish/x}K e^{-ish/x}f_x(s)-
K^\# f(s) \right)\ ds\\
& &=\int_0^t\  \left(e^{ish/x}K e^{-ish/x}(f_x(s)-f(s))
+\left( e^{ish/x}K e^{-ish/x} -  K^\# \right)f(s) \right) \ ds.
\nonumber
\eea
Hence,
\be\label{gw}
\|f_x(t)-f(t)\|\leq \|K\| \int_0^t\ \|f_x(s)-f(s)\|\ ds+{\cal F}(x,t_0)
\ee
where 
\be\label{inbp}
{\cal F}(x,t_0)=\sup_{0\leq t\leq t_0}\left\|\int_0^t\ 
\left( e^{ish/x}K e^{-ish/x}-K^\# \right)e^{-isK^\#}f\ ds\right\|.
\ee
Now, 
\bea
& & e^{ish/x}K e^{-ish/x}-K^\#= e^{ish/x}(K-K^\#) e^{-ish/x}\nonumber\\
& &\quad\quad\quad\quad=\sum_{j\neq k} e^{ish/x}P_jKP_k e^{-ish/x}=
\sum_{j\neq k}
e^{is(E_j-E_k)/x}P_jKP_k,
\eea
so that we can integrate (\ref{inbp}) by parts to obtain
\bea\label{agreg}
& &\int_0^t\ \left( e^{ish/x}K e^{-ish/x}-K^\# \right)e^{-isK^\#}f\ ds=\\
& &\sum_{j\neq k}\int_0^t\ \frac{x}{i(E_j-E_k)}
\frac{d}{ds}e^{is(E_j-E_k)/x} P_jKP_k e^{-isK^\#}f\ ds=\nonumber\\
& &\sum_{j\neq k}\frac{x}{i(E_j-E_k)} \left. e^{is(E_j-E_k)/x} P_jKP_k 
e^{-isK^\#}f\ \right|_0^t+\nonumber\\
& &\sum_{j\neq k}\int_0^t\ \frac{x}{(E_j-E_k)}
e^{is(E_j-E_k)/x} P_jKP_k K^\#e^{-isK^\#}f\ ds.\nonumber
\eea
Hence, using  $\|K^\#\|\leq \|K\|$, we can 
bound (\ref{agreg}) by
\be
\sum_{j\neq k} \frac{ x  \|K\|(2+t \|K\|)e^{ \|K\|t}}{|E_j-E_k|}\|f\|.
\ee
Thus, 
\be {\cal F}(x,t_0)\leq \max(2,t_0)(r^2-r)\frac{x(1+\|K\|)\|K\|e^{\|K\|t_0}}
{\inf_{j\neq k}|E_j-E_k|}. 
\ee
At this point we can invoke Gronwall's Lemma, the above estimate  
and (\ref{gw})  to finish the proof.
\ep\\

From these two Lemmas, we immediately get the 
\begin{prop}\label{weakprop}  Let $V(x)$, $x\in [0,x_0)$ and $R$ be 
bounded operators on a Banach  
space $\ban$ such that,  in the operator norm, 
$V(x)=V(0)+ x R + O(x^2)$, where $V(0)$ is an isometry admitting
the spectral decomposition $V(0)=\sum_{j=0}^r e^{-iE_j}P_j$ and 
let $h=\sum_{j=0}^r E_jP_j$. Then, 
for any $0\leq t\leq t_0$, if $x\ra 0$ in such a way that $t/x\in\N$, 
\be
V(0)^{-t/x}V(x)^{t/x}=e^{t{e^{ih}R}^\#}+O(x), \ \ \ \mbox{ in norm,}
\ee
where $K^\#=\sum_{j=0}^r P_jKP_j$, for any $K\in {\cal L}(\ban)$.
\end{prop}
{\bf Remarks: }\\
i) The operator in the exponent can be rewritten as
\be
e^{ih}R^\#=e^{ih}R^\#=(e^{ih}R)^\#=(Re^{ih})^\#. 
\ee
ii) The hypotheses are made on the isometry $V(0)$, not on the operator 
$h$.\\

\vspace{.3cm}

We can now derive our first results concerning the weak limit in the
Schr\"odinger picture. We do so
in the general setup described in {\bf H1}. We further assume:\\

\noindent
{\bf H2}: \\
The restriction $H_P(0)$ of $H(0)$ to $P\ban$ is diagonalizable
and reads 
\be
H_P(0)=\sum_{j=0}^{r} E_j P_j, \ \ \mbox{ with } \ \ \dim(P_j)\leq \infty, \ \ 
r \, \mbox{ finite }.
\ee
Moreover, the operator $Pe^{-i\tau H(0)}=Pe^{-i\tau H_P(0)}$
is an isometry on $P\ban$.\\
\vspace{.2cm}

Note that this implies $Pe^{-i\tau H_P(0)}$ is invertible and
\be
P=\sum_{j=0}^{r}P_j, \ \ \ E_j \in \R \ \ \ \forall 
j=0,\cdots, r,\ \ \ \mbox{ and }\ \ \
Pe^{-i\tau H(0)}=\sum_{j=0}^{r} e^{-i\tau E_j} P_j,
\ee
where the projectors $P_j$ are eigenprojectors of $Pe^{-i\tau H(0)}$ iff the
$e^{-i\tau E_j}$'s are distinct.
In case $\ban$ is a finite dimensional Hilbert space and $H(0)$ is self adjoint, 
{\bf H2} is automatically true.

\begin{prop} \label{1p}  
Let $H(\lambda)$ and $P$ on $\ban$ satisfy {\bf H1} and {\bf H2}. Further 
assume $\tau >0$ is such that the values are $\{e^{-i\tau E_j}\}_{j=0}^r$ 
are distinct. Then, for any $0\leq t<\infty$, 
\be
\lim_{\lambda\ra 0\atop t/\lambda^2\in\N}e^{i\tau t H(0)/\lambda^2}
\left[ Pe^{-i\tau H(\lambda)}P\right]^{t/\lambda^2}=e^{t\Gamma^w(\tau)} \ \ \mbox{on} \ \ P\ban\,
\ee
where
\be\label{gam}
\Gamma^w(\tau) = e^{i\tau H(0)}G(\tau)^\#
=-{\int_0^\tau ds \int_0^{s} dt  W e^{-it (H(0)-E_j)}
W}^\# ,
\ee
and $K^\#=\sum_j^r P_jKP_j$ for any $K\in {\cal L}(P\ban)$.
\end{prop}
{\bf Remarks:}\\
i) In case some values among $\{e^{-i\tau E_j}\}_{j=0}^r$ coincide, the result holds
whith the $P_j$'s replaced by $\Pi_j$'s, the spectral projectors of 
$Pe^{-i\tau H(0)}|_{P \ban}$.\\ 
ii) If, for any $j=0,\dots, r$, the reduced resolvents
$R_Q(E_j)=(H(0)-E_j)|_{Q\ban}^{-1}$
all exist, with $Q=\un -P$, then 
\bea\label{gamr}
\Gamma^w(\tau) &=& -\sum_{j=0}^r  P_j W R_Q(E_j) \left( R_Q(E_j)
  -R_Q(E_j)e^{-i\tau (H(0)-E_j)|_{Q\ban }} -i\tau \un  \right) W P_j   
\eea
iii) If $\ban$ is a Hilbert space, and $H(\lambda)$ is
self-adjoint with dim $P_j=1$, we can express $\Gamma^w$ in yet another way. 
We write $P_j=|\ffi_j\ket \bra \ffi_j |$ and introduce
$d \mu_j^W(E)$, $j=0,\cdots, r$, the spectral measures of the vectors
$W\ffi_j=QW\ffi_j$, with respect to $H(0)|_{ Q\ban}$. Then, if  
$\ \widehat{\cdot}\ $ denotes the Fourier transform,
\be\label{gamf}
\Gamma^w(\tau)=-\sum_{j=0}^r\int_0^\tau ds \int_0^{s} dt \ \widehat{\mu_j^W}(t) 
e^{itE_j}\ P_j.
\ee

\noindent
{\bf Proof of Proposition \ref{1p}:}\\
As we are to
work in $P\ban$, we will write $A_P$ for $PAP$ etc... Our assumption
on $\tau$ makes the eigenvalues of $e^{-i\tau H_P(0)}$ distinct so that 
the $P_j'$s are eigenprojectors of both $H_P(0)$ and $e^{_i\tau H_P(0)}$.  
Then, Lemma \ref{easypert} shows that $V(x):=P(e^{-i\tau (H(0)+\sqrt{x}W)}P$, $x=\lambda^2$, 
satisfies the hypotheses of Proposition \ref{weakprop} with
$h=\tau H_P(0)$, $R=G_P(\tau)$ and $\tau >0$ fixed. Hence the result, making use
of $e^{-i\tau H_P(0)}=e^{-i\tau H(0)}P$.
The last statement follows from (\ref{defg}). \ep

\vspace{.3cm}

We are now in  a position to state the existence
of a contraction  semi-group on $P\ban$ obtained by means of a weak
limit for our specific time dependent Hamiltonian model. The following is
a direct application of Proposition \ref{1p}.
\begin{cor}\label{c2}
Let $U(n,0)$ be defined on $\hil_0\otimes \hil $, where $\hil_0$ is separable, 
by (\ref{evol}, \ref{brick}, \ref{hamil}), let $P=\un\otimes |\Omega\ket\bra\Omega |$,
and let $\{E_j\}_{j=0, \cdots, r}$  be the eigenvalues of $h_0$
associated with eigenprojectors $\{P_j\}_{j=0,\cdots, r}$. Assume the
values  $\{e^{-i\tau E_j}\}_{j=0, \cdots, r}$ are distinct.
Then, for any fixed $0\leq t<\infty$, 
\be
\lim_{\lambda\ra 0\atop t/\lambda^2\in\N}\left[e^{i\tau t H(0)/\lambda^2}
PU(t/\lambda^2,0)P\right]=e^{t\Gamma^w(\tau)} \ \ \mbox{on} \ \ P\ban\,
\ee
where 
\be
\Gamma^w(\tau) = e^{i\tau H(0)}{G(\tau)}^{\#}=
-\int_0^\tau ds \int_0^{s} dt \sum_{j=0}^r\sum_{m=1}^n P_jV_m^*
e^{-it (h_0+\delta_m-E_j)}V_mP_j
\ee
generates a contraction semi-group and $\#$ corresponds to 
the  set of eigenprojectors $\{P_j\}_{j=0,\cdots, r}$.
\end{cor}
{\bf Remarks:} \\
0) The macroscopic time scale at which we observe the system is 
$T=\tau t/\lambda^2\ra \infty$.  \\
i) There are cases where $\Gamma^w(\tau)$ generates a group of isometries.\\
ii) Again, if the $e^{-i\tau E_j}$'s are not distinct, we have 
to take the spectral projectors of $e^{-i\tau h_0}$ instead of the 
$P_j$'s in the definition of the operation $\#$.\\
iii) Note that the effective dynamics commutes with $h_0$, so that
no transition between the eigenspaces of $h_0$ can take place.  However, 
if the $e^{-i\tau E_j}$'s are not distinct, transitions between different eigenspaces
of $h_0$ corresponding to the same eigenvalue of $e^{-i\tau h_0}$ are possible.
\\
iv) In case $h_0$ is non degenerate, $r=d$ and we can write $P_j=|x_j\ket \bra x_j|$, with
$x_j$ the eigenvector associated with $E_j$, and
\be
\Gamma^w(\tau) = -\sum_{j=0}^d \left(\sum_{k=0}^d\sum_{m=1}^n
|\bra x_k | V_m x_j\ket |^2 \int_0^\tau ds \int_0^s dt \, 
e^{-it(E_k-E_j+\delta_m)}\right) |x_j\ket\bra x_j |,
\ee
where the double integral equals
\be
 \int_0^\tau ds \int_0^s dt \, 
e^{-it\alpha}=\left\{\matrix{\tau^2/2& \alpha=0\cr 
\frac{1}{\alpha^2}(1-e^{-i\tau\alpha})-\frac{i}{\alpha}\tau 
 & \alpha\neq 0}\right.
\ee
\vspace{.3cm}

\noindent
{\bf Proof of Corollary \ref{c2}: }\\
By  Lemma \ref{red} above, 
\be
\left[e^{i\tau t H_P(0)/\lambda^2}
PU(t/\lambda^2,0)P\right]=e^{i\tau t H(0)/\lambda^2}\left[Pe^{-i\tau(H(0)+\lambda W)}P 
\right]^{t/\lambda^2},
\ee
where conditions {\bf H1} and {\bf H2} are met and Proposition \ref{1p} applies.  
The fact that $\Gamma^w(\tau)$ generates a contraction semigroup in 
that case stems from the a priori bound, uniform in $t, \tau, \lambda$,
\be
\|e^{i\tau t H(0)/\lambda^2}PU(t/\lambda^2,0)P\|\leq 1.
\ee
The expression for $\Gamma^w(\tau)$ comes from the explicit evaluation of
(\ref{gam}) in our model.
\ep

\subsection{Different Time Scales}

Looking at the dependence in $\tau$ of the result in Corollary \ref{c2}, we observe 
that we can obtain a different non-trivial effective evolution 
with our conventional weak limit approach, provided one further makes the
time scale $\tau\ra 0$ and, at the same time, increases
the parameter $t$ to $t/\tau^2$. This yields a macroscopic time
scale given by $T=t/(\tau \lambda^2)\ra \infty.$ We'll come back 
to this point also, when we deal with the Heisenberg evolution of
observables.\\

Using the first expression (\ref{defg}), one immediately gets
\be
\lim_{\tau\ra 0}\lim_{\lambda\ra 0\atop t/(\tau \lambda)^2\in\N}
\left[e^{i\tau t H(0)/(\tau\lambda)^2}
PU(t/(\tau \lambda)^2,0)P\right]=\lim_{\tau\ra 0}e^{t\Gamma^w(\tau)/\tau^2}
\equiv e^{t\Gamma^{1}},
\ee
where 
\be
\Gamma^1={\Gamma_0}^\#=\sum_{j=0}^r P_j \Gamma_0 P_j, \ \ \ 
\Gamma_0=-\frac12 \sum_{i=1}^nV_iV_i^*.
\ee
Note that under the hypotheses of Corollary
\ref{c2}, the spectral projectors of
$h_0$ and $e^{-i\tau h_0}$ coincide when $\tau\ra 0$.\\

This calls for a redefinition of the scaling, right from
the beginning of the calculation, in order to arrive at the 
same result, without resorting to iterated limits, as above.
This is at this point that we need to consider the dependence 
in $\tau$ of the previous steps.

\vspace{.3cm}

We state below is our main theorem regarding this issue in the general
Banach space framework under hypotheses {\bf H1} and {\bf H2}. Actually, the application 
above is a consequence of the theorem to come.  The study at positive temperature 
in Heisenberg picture of the forthcoming Sections will rely on this result as well.

\begin{thm}\label{theo31}
Suppose Hypotheses {\bf H1} and {\bf H2} hold true and further assume the spectral projectors
$P_j$, $j=0,\cdots, r$, of $e^{-i\tau H_P(0)}$ coincide with those of $H_P(0)$ 
on $P\ban$. Set  $K^\#=\sum_{j=0}^r P_j K P_j$, for $K\in {\cal L}(\ban)$.\\

A) Then, for any 
$0<t_0<\infty$, there exists $0<c<\infty$ such that for any $0\leq t \leq t_0$,
the following estimate holds in the limit $\lambda^2\tau \ra 0$, $\lambda^2\tau^2\ra 0$, 
and $t/(\lambda\tau)^2\in \N$:
\be
\left\|e^{iH(0)t/(\lambda^2 \tau)}\left[Pe^{-i\tau (H(0)+\lambda W)}
P\right]^{t/(\lambda \tau)^2}-e^{t \, e^{i\tau H(0)}G_P(\tau)^{\#}/\tau^2}\right\|
\leq c(\lambda^2\tau^2 +\lambda^2\tau).
\ee

B)  Then, for any 
$0<t_0<\infty$, there exists $0<c<\infty$ such that for any $0\leq t \leq t_0$,
the following estimate holds in the limit $\lambda^2\tau \ra 0$, $\tau\ra 0$, 
and $t/(\lambda\tau)^2\in \N$:
\be
\left\|e^{iH(0)t/(\lambda^2 \tau)}\left[Pe^{-i\tau (H(0)+\lambda W)}
P\right]^{t/(\lambda \tau)^2}-e^{-t(W^2)^{\#}/2}\right\|
\leq c(\tau +\lambda^2\tau).
\ee

\end{thm}
{\bf Remarks:} \\
0) If $\tau$ is small enough, the spectral projectors of $e^{-i\tau H_P(0)}$ and $H_P(0)$ 
on $P\ban$ coincide. \\
i) If $\tau$ is fixed, part A) of the Theorem coincides with Proposition \ref{1p} with 
$\tilde t := t/\tau^2$ in place of $t$.\\

\noindent
{\bf Proof:} We only need to consider the case $\tau>0$ small, 
where  Remark 0) applies. We proceed in two steps, using Lemmas \ref{step1} and \ref{step2} in sequence.
Let $x=\lambda^2\tau^2$. The expansions provided in Lemma \ref{easypert} yield 
\be
Pe^{-i\tau (H(0)+\lambda W)}P=e^{-i\tau H_P(0)}+xG_P(\tau)/\tau^2 + O(x^2),
\ee
with $G_P(\tau)/\tau^2=O(1)$ and reminder uniform in $\tau\ra 0$. Hence,
\be
\| Pe^{-i\tau (H(0)+\lambda W)}P\| = 1 + O(x), \ \ \ \mbox{ uniformly in } \ \ \tau.
\ee
As $e^{-i\tau H_P(0)}=\sum_j^re^{-i\tau E_j}P_j$, with 
$P_j$ independent of $\tau$, the operator $J(\tau)$ defined in (\ref{defj}) reads
\be
J(\tau)=\sum_{j,k=0}^r P_j\frac{G_P(\tau)}{\tau^2} P_k\alpha_{jk}(\tau),  \ \ \mbox{ where } \ \
 \alpha_{jk}(\tau)=\left\{
\matrix{\frac{\tau(E_j-E-k)}{e^{-i\tau E_j}-e^{-i\tau E_k}} & j\neq k \cr 
ie^{i\tau E_j}& j = k.}\right.
\ee
Hence,
\be\label{jb}
\alpha_{jk}(\tau)=i+O(\tau), \ \
\frac{G_P(\tau)}{\tau^2}=-\frac{{W^2}_P}{2}+O(\tau) \ \  
\mbox{ and } \ \ J(\tau)=O(1) \ \ \mbox{ as }\ \ \tau \ra 0.
\ee 
Now, using (\ref{ff}) with coupling constant $x/\tau$ (and the first remark following 
Lemma \ref{easypert}), we can write for $x/\tau$ small, uniformly in $\tau$,
\bea
&&e^{-i\tau ( H_P(0)+\frac{x}{\tau} J(\tau))}=\\
&&\hspace{.5cm}e^{-i\tau H_P(0)}+\frac{x}{\tau}
\left( -ie^{-i\tau H_P(0)}\int_0^\tau \e^{is H_P(0)}J(\tau)e^{-is H_P(0)}  
ds\right)
+ O((x/\tau)^2\tau^2) =\nonumber\\ 
&&\hspace{.5cm}e^{-i\tau H_P(0)}+x \left( -ie^{-i\tau H_P(0)}\int_0^1 
\e^{is \tau H_P(0)}J(\tau)e^{-is \tau H_P(0)} 
 ds\right)
+O(x^2),\nonumber
\eea
where the operator in the bracket above is $O(1)$ as $\tau\ra 0$. Hence 
\be
\|e^{-i\tau ( H_P(0)+\frac{x}{\tau} J(\tau))}\|=1+O(x), \ \ \ \mbox{ uniformly in } \ \ \tau.
\ee
Thus, we  apply  Lemma \ref{step1}, to get
\be
\left\| \left[Pe^{-i\tau (H(0)+\lambda W)}P\right]^{\frac{t}{x}}-
e^{-i\frac{t}{x}(\tau  H_P(0)+x J(\tau))}\right\|=O(x^2), 
\ee
 as $x\ra 0$, and $\frac{x}{\tau} \ra 0$, with a remainder uniform in $\tau$. 

We now turn 
to the second step. We can write
\be
e^{-i\frac{t}{x}(\tau  H_P(0)+x J(\tau))}=
e^{-i\frac{t}{\lambda^2\tau}(H_P(0)+\lambda^2\tau J(\tau))}\equiv 
e^{-i\frac{t}{y}(H_P(0)+y J(\tau))}\ \ \ \mbox{ with } y=\lambda^2\tau.
\ee
Therefore, by Lemma \ref{step2} and the last statment of (\ref{jb}),
\be
e^{i\frac{t}{y}H_P(0)}e^{-i\frac{t}{y}(H_P(0)+y J(\tau))}-e^{-it J^\#(\tau))}=O(y),
\ee
uniformly in $\tau$. Hence, for any given $t_0$,  we get the existence of a 
constant $0<c<\infty$, uniform in $\tau$, such that for all $0<t\leq t_0<\infty$,
\be
\left\|e^{i\frac{t}{\lambda^2\tau }H_P(0)} 
\left[Pe^{-i\tau (H(0)+\lambda W)}P\right]^{\frac{t}{(\lambda \tau)^2}}- 
e^{-it J^\#(\tau))}\right\|\leq c (\lambda^2\tau+\lambda^2 \tau^2),
\ee
as $\lambda^2\tau$ and $\lambda^2 \tau^2$ go to zero in such a way that 
$t/(\lambda \tau)^2\in \N$, which is part A) of the Theorem.
Part B) follows from the first statements in (\ref{jb}) and of the fact 
that the projectors $P_j$'s are independent of $t$. \ep\\

As a direct Corollary, we get,

\begin{cor}\label{3.5} Let $U(n,0)$ be defined on $\hil_0\otimes \hil $, 
$\hil_0$ a separable Hilbert space, by (\ref{evol},
\ref{brick}, \ref{hamil}), let $P=\un\otimes |\Omega\ket\bra\Omega |$,
and let $\{E_j\}_{j=0, \cdots, r}$ be the eigenvalues of $h_0$ 
associated with eigenprojectors $\{P_j\}_{j=0,\cdots, r}$. 
Then, for any $0\leq t \leq t_0$,
\be
\lim_{\tau\ra 0, \lambda^2\tau\ra 0\atop
 t/(\tau \lambda)^2\in\N}
\left[e^{i\tau t H(0)/(\tau\lambda)^2}
PU(t/(\tau \lambda)^2,0)P\right]=e^{t\Gamma_{0}^\#},
\ee
where ${\Gamma_0}^\#=\sum_{j=0}^r P_j \Gamma_0 P_j,$ and 
$\Gamma_0=-\frac12 \sum_{i=1}^nV_iV_i^*.$
\end{cor}

\section{Heisenberg representation for non-zero temperature}\label{sechei}
\setcounter{equation}{0}

From now on, we stick to our model Hamiltonian characterized by hypothesis 
{\bf H0}.  We first express the evolution at positive temperature 
of observables $B$ of the small system (\ref{bkt}) after 
$k$ repeated interactions as the action
of the $k$-th power of an operator ${\cal U}_\beta(\lambda, \tau)$ on $\hil_0$. 
This reflects the Markovian nature of our model.

This is done in Proposition
\ref{bkuk}. This allows us to apply Theorem \ref{theo31} again to 
compute the weak limit in Theorem \ref{mtb}. Let us mention here
already that we perform a complete
analysis of the special case where both the small system and the individual
spins of the chain live in $\C^2$ in the last Section of the paper.\\

Let us define the equilibrium state $\omega(\beta)_N$ of a chain 
of $N$ spins at inverse temperature $\beta$ by a tensor product 
of individual diagonal density matrices of the form
\be
r(\beta)=\frac{1}{1+\sum_{j=1}^ne^{-\beta \delta_j}}
\pmatrix{1 & 0 & \cdots & 0
\cr 0 & e^{-\beta \delta_1}&  \cdots & 0
\cr \vdots &  & \ddots & \vdots
\cr 0 &\cdots  &0 & e^{-\beta \delta_{n}}  }=
\frac{e^{-\beta \delta a^*a}}{{\cal Z}(\beta)},
\ee
in the basis $\{\omega , x_1, \cdots, x_n \}$ of $\C_j^{n+1}$, i.e.
\be
\omega(\beta)_N=r(\beta)\otimes r(\beta)\otimes \cdots \otimes r(\beta).
\ee

The individual density matrices $r(\beta)$ are defined by Gibbs 
prescription for the Hamiltonians at each site 
\be
\sum_{i=1}^n\delta_ia_i^*a_i
\ee
corresponding to our model (\ref{hamfield})

Our spin chain is of finite length $N$, but, as we will see below,
only the first $k$ spins matter to study the time evolution up to
time $k$. This will allow us to take the thermodynamical limit
by hand.
If $\rho$ is any state on $\C^{d+1}$, the initial state of the
small system plus spin chain is 
$\rho\otimes \omega(\beta)_N$. We shall study the Heisenberg
evolution of observables of the form $B\otimes \un_{\cal H}$, where 
$B\in M_{d+1}(\C)$, defined  by
\be\label{bkt}
B_\beta(k,\lambda,\tau)=\tr_{\cal H}((\un\otimes\omega_N(\beta)) \,\, U(k,0)^{-1}
(B\otimes \un_{\cal H})U(k, 0)),
\ee
where, for any $A\in {\cal L}(\hil_0\otimes \hil)$,
\be\label{partra}
\tr_{\cal H}(A)=\left(\sum_{S}\bra x_i\otimes x_S |\, A 
\, x_j\otimes x_S\ket 
\right)_{i,j\in\{0,\cdots, d\}}\ \ \  \mbox{ with } \ \ \  x_0=\omega, 
\ee
 denotes the partial trace taken on the spin variables only.
Hence, the expectation in the state $\rho$ of the observable 
$B$ after $k$ interactions
over a time interval of length $k\tau$ with the chain at 
inverse temperature $\beta$ is given by
\be
\bra B(k, \beta)\ket_\rho=\tr_{\C^{d+1}}(\rho B_\beta(k,\lambda,\tau)).
\ee
{\bf Remark:}\\
In case $\hil_0$ is infinite dimensional, the definitions (\ref{bkt}) 
and (\ref{partra}) hold, {\it mutatis mutandis}. For instance, consider 
$B\in{\cal L}(\hil_0)$ in (\ref{bkt}), where (\ref{partra}) should be
read as 
\be
\tr_{\cal H}(A)=\sum_{S}\bra \un_{\hil_0}\otimes  x_S |\, A 
\, \un_{\hil_0}\otimes|x_S \ket,
\ee
with a slight abuse of notations.

\subsection{Markov Properties}

Recall that
\be
U(k,0)^{-1}(B\otimes \un_{\cal H})U(k, 0)=
U_1^* U_{2}^*\cdots U_k^* (B\otimes \un_{\cal H})U_k U_{k-1}\cdots U_1,
\ee
where $U_j$ is non-trivial on $\C^{d+1}\otimes \C_j^{n+1}$ only. 

Let us specify a bit more the partial trace operator $\tr_{\cal H}
((\un\otimes\omega_N(\beta))\,\, A)$, where $A$ is an operator
on $\C^{d+1}\otimes \Pi_{j=1}^N\C_j^{n+1}$.

\begin{lem}
Let us denote the matrix elements of $A$ as follows
\be
A^{i,j}_{S, S'}=\bra x_i\otimes X_S |A \,\, 
x_j\otimes X_{S'}\ket,
\ee
where $i,j$ belong to  $\{0,\cdots, d\}$, and
$S$, $S'$ run over subsets of $\left\{\{1, \cdots, N\}\times \{1,\cdots,n\}
\right\}^N$ as in (\ref{defs}) Then
\be\label{4.9}
\tr_{\cal H}
((\un\otimes\omega_N(\beta))\,\, A)_{i,j}=
\sum_S\frac{e^{-\beta\sum_{l=1}^{n}\delta_l |S|_l}}
{(1+\sum_{l=1}^{n}e^{-\beta\delta_l})^N}A_{S,S}^{i,j}
\ee
where, for
\be
S=\{ (k_1, i_1), (k_2,i_2), \cdots, (k_m, i_m) \} \subset 
(\N\times \{1,2, \cdots , n\})^m 
\ee 
 with all  $1\leq k_j \leq N$ distinct and $m=0,\cdots, N$,
\be
|S|_l=\#\{k_r\ \  \mbox{s.t.}\ \ i_r=l\}.
\ee
\end{lem}
{\bf Proof:} Follows directly from
\be
\omega_N(\beta)X_S=\frac{\Pi_{r=1}^me^{-\beta \delta_{i_r}}}{(1+\sum_j
e^{-\delta_j\beta})^N}X_S=
\frac{e^{-\beta \sum_{l=1}^n\delta_l |S|_l}}{(1+\sum_j
e^{-\delta_j\beta})^N}X_S.
\ee
\hfill\ep

We now further compute the action of $U(k,0)$ given by the product of
$U_j's$. 
Let us denote the vectors $\omega\otimes X_S$ and $x_j\otimes X_S$ by 
$n_0\otimes |n_1, n_2, \cdots , n_N\ket
\equiv n_0\otimes |\vec n \ket $, where $n_0\in\{0,1,\cdots d\}$,
and  $n_j\in \{0,1, \cdots n\}$, for any $j=1,
\cdots N$, with $\omega \simeq  0$ and $x_k \simeq k$ and 
$X_{\{(1,n_1), \cdots, (N,n_N)\}}\simeq |\vec n \ket$.

Recall that
\be
U_j=e^{-i\tau\widehat{H}_j}e^{-i\tau \widetilde{H}_j},
\ee
where $e^{-i\tau\widehat{H}_j}$ is diagonal. More precisely, with 
the convention $\delta_0=0$,
\be
e^{-i\tau\widehat{H}_j}n_0\otimes |n_1, n_2, \cdots , n_N\ket=
e^{-i\tau\sum_{k=1\atop k\neq j}^N\delta_{n_k}}n_0\otimes |n_1, n_2, \cdots , n_N\ket.
\ee
\begin{lem}\label{matel}
 Denoting the $k$-independent matrix
elements of $e^{-i\tau\widetilde{H}_k}|_{\C^{d+1}\otimes \C_k^{n+1}}=
\tilde{U}_k|_{\C^{d+1}\otimes \C_k^{n+1}}$ by 
\be
U^{n,n'}_{m,m'}=\bra n\otimes m | \tilde{U}_k \, n'\otimes m'\ket,  
\ee
we have for any
$N\geq k$
\bea
&& U_kU_{k-1}\cdots U_2U_1 \,\, n_0\otimes |n_1,\cdots, n_N\ket=\\
&&\sum_{\vec{m_0}\in \{0,\cdots, d\}^{k} \atop \vec m\in \{0,\cdots, n\}^{k} }
e^{-i\tau \ffi(\vec m,\vec n)}U_{m_k,n_k}^{m_0^k,m_0^{k-1}}
\cdots U_{m_2,n_2}^{m_0^2,m_0^1} U_{m_1,n_1}^{m_0^1,n_0}\,\,  
m_0^k\otimes |m_1, m_2,\cdots, m_k, n_{k+1}, \cdots, n_N\ket,\nonumber
\eea
where 
\be\label{phase}
\ffi(\vec m,\vec n)= \sum_{j=1}^{k} \left(\sum_{j < l \leq N}\delta_{n_l}+
\sum_{l < j}
\delta_{m_l}\right)
\ee
\end{lem}
{\bf Proof:} Consequence of the iteration of formulae of the
type
\be
U_1 \,\, n_0\otimes |n_1,\cdots, n_N\ket=\sum_{m_0^1=0,1, \cdots , d \atop 
m_1=0,1, \cdots, n}
e^{-i\tau \sum_{j>1}\delta_{n_j}}
U_{m_1,n_1}^{m_0^1,n_0}\,\, m_0^1\otimes |m_1, n_2, n_3, \cdots, n_N\ket.
\ee
\hfill \ep

A consequence of these formulae is that we 
can consider  spin chains consisting in $k$ spins only:
\begin{lem} For any $N\geq k$,
\bea
&&\tr_{\cal H}(\un\otimes\omega_N(\beta) \,\, U_1^* U_{2}^*\cdots U_k^*
(B\otimes \un_{\cal H})U_k U_{k-1}\cdots U_1)=\nonumber\\
&&\hspace{4cm}\tr_{\cal H}(\un\otimes\omega_k(\beta) \,\, 
U_1^* U_{2}^*\cdots U_k^*
(B\otimes \un_{\cal H})U_k U_{k-1}\cdots U_1)
\eea
\end{lem}
{\bf Proof:} Obvious from the tensor product structure of 
$\omega_N(\beta)$. \hfill \ep

\vspace{.3cm}

To proceed, let us adopt the following block notation
\be\label{4.20}
U=e^{-i\tau (H(0)+\lambda W)}=\pmatrix{U_{0,0} & U_{0,1}& \cdots & U_{0,n}\cr 
U_{1,0} & U_{1,1}& \cdots & U_{1,n}\cr \vdots &\vdots &\ddots & \vdots
\cr U_{n,0} & U_{n,1}& \cdots & U_{n,n} } 
\ee where  
\be
U_{m,m'}=\pmatrix{U_{m,m'}^{0,0} & U_{m,m'}^{0,1}& \cdots & U_{m,m'}^{0,d}\cr 
U_{m,m'}^{1,0} & U_{m,m'}^{1,1}& \cdots & U_{m,m'}^{1,d}\cr 
\vdots &\vdots &\ddots & \vdots
\cr U_{m,m'}^{d,0} & U_{m,m'}^{d,1}& \cdots & U_{m,m'}^{d,d} }.
\ee
In terms of the notations of the previous Section,  
\be
U=\pmatrix{PUP & PUQ \cr QUP& QUQ},
\ee 
we have the identifications
\bea\label{identif}
& &PUP\simeq U_{0,0}, \ \ \ \ \  QUQ\simeq \pmatrix{ U_{1,1}& \cdots & U_{1,n}\cr 
\vdots &\ddots & \vdots
\cr  U_{n,1}& \cdots & U_{n,n} },\nonumber \\ 
& &PUQ\simeq \pmatrix{ U_{0,1}& \cdots & U_{0,n}}, \ \ \ \ 
QUP\simeq \pmatrix{ U_{1, 0}& \cdots & U_{n, 0}}^T.
\eea

Let us finally denote the inverse of $U=(U^{n,n'}_{m,m'})$
by 
\be
V=({V}^{n,n'}_{m,m'})=U^{-1}=({U^{-1}}^{n,n'}_{m,m'})\in M_{(1+d)(1+n)}(\C),
\ee
so that we have for any $m$ and $n$
\be
U_{n,m}^*=V_{m,n}\in M_{1+d}(\C).
\ee
With these notations, we have
\begin{lem} The  matrix elements of 
$U(k,0)^{-1}\,\, (B\otimes \un_{\cal H}) \,\, U(k, 0)$
in the orthonormal basis $\{n_0\otimes |n_1,\cdots,n_k\ket\}=\{n_0\otimes |\vec{n}\ket\}$ read
\bea
& &\bra \tilde{n_0}\otimes \vec{\tilde{n}} |(U_k\cdots U_1)^* B\otimes 
\un_{\cal H}\,(U_k\cdots U_1)
\,\,n_0\otimes\vec{n}\ket=\\
& &\quad \quad  e^{-i\tau(\ffi(0,\vec n)-\ffi(0 ,\vec{\tilde{n}}))}
\sum_{\vec m\in\{0,\cdots, n\}^k} (V_{\tilde n_1, m_1}\cdots V_{\tilde n_k,m_k}BU_{m_k,n_k}\cdots U_{m_1, n_1})^{\tilde n_0, n_0}
\nonumber
\eea
\end{lem}

{\bf Proof:} Expand the products and make use of Lemma \ref{matel} and 
(\ref{phase}). \hfill \ep\\

The above Lemmas and (\ref{bkt}) lead us to the study of the matrix in $M_{d+1}(\C)$
\be
B_\beta(k,\lambda,\tau)=\sum_{\vec{n}=(n_1,\cdots, n_k)\atop \vec m=(m_1,\cdots m_k)}
\frac{e^{-\beta\sum_{l=0}^{n}\delta_l|\vec n|_l}}{(1+\sum_{j=1}^n e^{-\delta_{j}\beta})^k}V_{ n_1, m_1}\cdots 
V_{n_k,m_k}BU_{m_k,n_k}\cdots U_{m_1, n_1}
\ee 
in various  limiting cases as $\lambda$ and/or $\tau$ go to zero, with the notation 
\be
|\vec n|_l=\sharp\{n_r \ \ \mbox{s.t.}\ \ n_r=l\}= |S|_l.
\ee 

We introduce operators on the Hilbert space $M_{d+1}(\C)$ 
equipped with the scalar product $\bra A | B\ket =\tr (A^* B)$, 
for any $A, B\in M_{d+1}(\C)$ by 
\be\label{calu}
{\cal U}_{m,m'}(A):=V_{m',m}\, A\, U_{m,m'}, \,\,\,\,\,\, (m,m')\in 
\{0,1,\cdots, n\}^2.
\ee
These operators are linear and one has with respect to the above scalar product,
\be
{\cal U}_{m,m'}^*(\cdot)=(V_{m',m}\, \cdot \, U_{m,m'})^*=  U_{m,m'}\, 
\cdot \, V_{m',m}.
\ee
The composition of such operators will be denoted as follows
\be
{\cal U}_{m',n'}\, {\cal U}_{m,n}(A)=V_{n', m'} V_{n,m}\, A\, U_{m,n} U_{m',n'}.
\ee

We are now in a position to express the Markovian nature of the evolution of 
our observables:
\begin{prop}\label{bkuk}
In terms of the operators defined above, we can write 
\bea
B_\beta(k,\lambda,\tau)=&\frac{1}{(1+\sum_{j=1}^ne^{-\delta_j\beta})^k}&
\left({\cal U}_{0,0}+e^{-\beta\delta_1}{\cal U}_{0,1}+\cdots +
e^{-\beta\delta_n}{\cal U}_{0,n}\right.\nonumber\\
&&+{\cal U}_{1,0}+e^{-\beta\delta_1}{\cal U}_{1,1}+\cdots +
e^{-\beta\delta_n}{\cal U}_{1,n}\nonumber\\
&&+\left.
{\cal U}_{n,0} +e^{-\beta\delta_1}{\cal U}_{n,1}+
\cdots + e^{-\beta\delta_n}{\cal U}_{n,n} \right)^k(B)\nonumber\\
&&\equiv {\cal U}_\beta(\lambda,\tau)^k(B).
\eea
\end{prop}
{\bf Proof:} By definition of ${\cal U}_{m,n}$ we have 
\be
B_\beta(k,\lambda,\tau)=\sum_{\vec{n}=(n_1,\cdots, n_k)\atop \vec m=(m_1,\cdots m_k)}
\frac{e^{-\beta\sum_{l=1}^{k}\delta_{n_l}}}{(1+\sum_{j=1}^ne^{-\delta_j\beta})^k}{\cal U}_{m_1,n_1}\cdots 
{\cal U}_{m_k, n_k}(B).
\ee
Furthermore introducing $
{\cal Y}_{m,n}=e^{-\delta_n\beta  }{\cal U}_{m,n}$, we get
\be
B_\beta(k,\lambda,\tau)=\frac{1}{(1+\sum_{j=1}^ne^{-\delta_j\beta})^k}\sum_{\vec{n}=
(n_1,\cdots, n_k)\atop \vec m=(m_1,\cdots, m_k)}
{\cal Y}_{m_1, n_1}\cdots 
{\cal Y}_{m_k, n_k}(B).
\ee
There are $(n+1)^2$ distinct operators ${\cal Y}_{m,m'}$ in that expression, 
and the set of vectors $\vec n, \vec m$ in the
sum yields all different ways of composing  $k$ of them. Therefore
\be
B_\beta(k,\lambda,\tau)=\frac{1}{(1+\sum_{j=1}^ne^{-\delta_j\beta})^k}
\left({\cal Y}_{0,0}+\cdots +{\cal Y}_{0,n}+\cdots +{\cal Y}_{n,0} +
\cdots + {\cal Y}_{n,n} \right)^k(B).
\ee
\hfill \ep\\
{\bf Remark:} \\
The formula of Proposition \ref{bkuk} holds if $\hil_0$ is a separable
Hilbert space, provided the decomposition of operators $A$ in (\ref{4.20})
is interpreted as $A_{pq}\in {\cal L}(\hil_0)$, $q,p\in \{1,\cdots, n\}$, with 
\be
A_{pq}=\un_{\hil_0}\otimes |p\ket\bra p |\, A\, \un_{\hil_0}\otimes |q\ket\bra q | ,
\ee
and the identification $\hil_0\otimes \C |q \ket\simeq \hil_0$, for all $q$.

\subsection{Weak Limit in the Heisenberg Picture}

The $\lambda$-dependence in $B_\beta(k,\lambda,\tau)$ comes from the definition 
\be\label{deful}
U=U_\tau(\lambda)=e^{-i\tau(H(0)+\lambda W)},
\ee
which implies that the ${\cal U}_{n,m}$'s depend on $\lambda$
as well, in an analytic fashion, and will be denoted 
${\cal U}_{n,m}(\lambda)$.
Expliciting the $\lambda$ dependence in $B_\beta(k,\lambda,\tau)$, the weak limit 
corresponds to taking $k=t/\lambda^2$ and computing the behavior of
$B_\beta( t/ \lambda^2,\lambda,\tau)$, as $\lambda\ra 0$ (keeping $\tau$ fixed). 
We shall use the same strategy as in the 
previous Section and Lemma \ref{pertuu} to identify the weak limit by
means of perturbation theory. We shall also eventually consider the
possibility of letting $\tau\ra 0$, therefore we explicit the behavior in
$\tau$ of the expansions below. \\

Consequently, with (\ref{identif}) and Corollary \ref{pertuu}, we get
\begin{lem}\label{4.5} Let $U$ be given by (\ref{deful}), with $H(0)$, $W$ 
self adjoint and satisfying {\bf H1}, and further assume $H(0)$ is 
diagonal with respect to the basis (\ref{ordb}). 
If ${\cal U}_{m,m'}(\lambda)$ is defined by (\ref{calu}) 
As $\lambda \ra 0$, we get the expansions
\bea
&&{\cal U}_{0,0}(\lambda)={\cal U}_{0,0}(0)+\lambda^2{\cal U}_{0,0}^{(2)}
+O(\lambda^4\tau^4)\\
&&{\cal U}_{m,m'}(\lambda)={\cal U}_{m,m'}(0)+\lambda^2{\cal U}_{m,m'}^{(2)}
+O(\lambda^4\tau^4), \ \ \ m, m' \geq 1 \\
&&{\cal U}_{0,m}(\lambda)=\lambda^2{\cal U}_{0,m}^{(1)}+O(\lambda^4\tau^4), 
\,\,\,\,\, m\geq 1\\
&&{\cal U}_{m,0}(\lambda)=\lambda^2{\cal U}_{m,0}^{(1)}+O(\lambda^4\tau^4), 
\,\,\,\,\, m\geq 1
\eea
where, for all $0\leq m, m' \leq n$
\bea
&&{\cal U}_{m,m'}(0)(B)=\delta_{m,m'}e^{i\tau H_{m,m}(0)}\, 
B \,e^{-i\tau H_{m,m}(0)}, \ \ \  \\
&&{\cal U}_{m,m'}^{(2)}(B)=\delta_{m,m'}(G_{m,m}(-\tau)Be^{-i\tau H_{m,m}(0)}+
e^{i\tau H_{m,m}(0)}BG_{m,m}(\tau)),
\eea
and, for all $1\leq m$,
\bea
&&{\cal U}_{0,m}^{(1)}(B)=F_{m,0}(-\tau)BF_{0,m}(\tau),
\\
&&{\cal U}_{m,0}^{(1)}(B)=F_{0,m}(-\tau)BF_{m,0}(\tau).
\eea
\end{lem}

This Lemma allows us to perform the analysis of
the operator defined in Proposition \ref{bkuk}
\be\label{ubeta}
{\cal U}_\beta(\lambda,\tau)={\cal Z}(\beta)^{-1}
\sum_{0\leq m \leq n\atop 0\leq l\leq n}{\cal U}_{l,m}(\lambda)
e^{-\delta_{m}\beta}, \ \ \ \mbox{as}\ \ \ \lambda \ra 0,
\ee 
with the convention $\delta_0=0$ and  
${\cal Z}(\beta)=\sum_{j=0}^ne^{-\delta_j\beta}$. 
Recall that
\be
B_\beta(k,\lambda,\tau)={\cal U}_\beta(\lambda,\tau)^k(B).
\ee
Moreover, using the fact, see (\ref{H(0)}),
\be
H_{m,m}(0)=H_{0,0}(0)+\delta_m \simeq h_0+\delta_m,
\ee
we get for all $0\leq m\leq n$
\be
{\cal U}_{m,m}(0)(B)={\cal U}_{0,0}(0)(B)\simeq e^{i\tau h_0} B e^{-i\tau h_0}
=e^{i\tau [h_0,\cdot]}(B).
\ee
We have thus shown the 
\begin{lem} Assume the hypotheses of Lemma \ref{4.5}. Then  
\bea \label{tbeta}
{\cal U}_\beta(\lambda, \tau)&=&{\cal U}_{0,0}(0)
+\frac{\lambda^2}{ {\cal Z}(\beta)}\left[\sum_{m=1}^n\left\{e^{-\beta\delta_m}
\left({\cal U}_{0,m}^{(1)}+{\cal U}_{m,m}^{(2)}\right)
+{\cal U}_{m,0}^{(1)}
\right\} +{\cal U}_{0,0}^{(2)}\right]+O(\lambda^4\tau^4)\nonumber\\
 &\equiv&{\cal U}_{0,0}(0)
+\lambda^2 {\cal Z}(\beta)^{-1}T_\beta +O(\lambda^4\tau^4),
\eea
with $T_\beta=T_\beta(\tau)=O(\tau^2).$
\end{lem}
The above operator enjoys the following symmetry property
\begin{lem}\label{symt} For any $B\in M_{d+1}(\C)$,
\be
\tr (BT_\beta(B^*))=\overline{\tr (B^*T_\beta(B))}.
\ee
\end{lem}
{\bf Proof:} Due to 
\bea
F_{n,m}(-\tau)&=&F_{m,n}(\tau)^*, \,\,\, m\neq n\\
G_{n,n}(-\tau)&=&G_{n,n}(\tau)^*
\eea
and to the structure of $T_\beta$, the result will be
proven once we show that for all $A,B,C\in M_{d+1}(\C)$
\be
\overline{\tr (B^*ABC+B^*C^*BA^*)}=\tr (BAB^*C+BC^*B^*A^*).
\ee
But this follows from  $\tr B =\tr B^T$, where $\cdot^T$ denotes the
transpose, and from the cyclicity of the trace again.\hfill\ep
\vspace{.3cm}

Recall also the property 
\be\label{tbu}
{\cal U}_\beta(\lambda, \tau)(\un)=\un \ \ \ \Rightarrow \ \ \ 
T_\beta(\un)=0,
\ee
and the fact that in case the spectrum $\{E_j\}_{j =0,\cdots, d}$ of $h_0$ 
is non-degenerate and $\{|x_j\ket\}_{j =0,\cdots, d}$ denotes the corresponding eigenvectors,
the unitary ${\cal U}_{0,0}(0)$ has degenerate spectrum:
\be\label{deg}
{\cal U}_{0,0}(0)(|x_j\ket\bra x_k| )= e^{i\tau(E_j-E_k)} \ |x_j\ket\bra x_k|, \ \ \
\forall  \ 0\leq j, k \leq d .
\ee
That is, 
$\sigma({\cal U}_{0,0}(0))=\{ e^{i\tau(E_j-E_k)} \}_{0\leq j, k \leq d}$,
so that $1$ is $d+1$ times degenerate at least. 

\vspace{.3cm}

We are in the same position as in the proof of Proposition
(\ref{1p}).
Therefore,
we can compute the weak limit from Proposition {\ref{weakprop}} immediately to get the following
\begin{thm}\label{mtb} Let ${\cal U}_\beta(\lambda,\tau)$ be given by (\ref{ubeta}), and
${\cal U}_{0,0}(0)$,  $T_\beta$ by (\ref{tbeta}). 
Let $\{e^{i\tau \Delta_l}\}_{l=1,\cdots, r}$
be the set of distinct eigenvalues of ${\cal U}_{0,0}(0)$
and denote by $P_l$ the corresponding orthogonal projectors.
Then
\bea
&&\lim_{\lambda\ra 0\atop t/\lambda^2\in\N }{\cal U}_{0,0}(0)^{-t/\lambda^2}
B_\beta( t/\lambda^2,\lambda,\tau)=\\
&&\quad \quad\quad\quad\quad \lim_{\lambda\ra 0\atop t/\lambda^2\in\N}
{\cal U}_{0,0}(0)^{-t/\lambda^2}
{\cal U}_\beta(\lambda,\tau)^{t/\lambda^2}(B)=e^{t\Gamma^w_\beta}(B),\nonumber
\eea
were
\be
\Gamma^w_\beta(B)=\frac{1}{{\cal Z}(\beta)}\left({\cal  U}_{0,0}(0)^{-1}\ T_\beta
\right)^\#(B),
\ee
with $\#$ corresponding to the set of projectors
$\{P_l\}_{l=1,\cdots, r }$.
\end{thm}
{\bf Remarks:}\\
0) In order to make the generator $\Gamma^w_\beta$ completely explicit, one needs 
to analyse the properties $T_\beta$, i.e. of the operators $V_j$ 
defining the coupling, within the eigenspaces of ${\cal U}_{0,0}(0)$. 
A non trivial example is worked out in Section 6, see Proposition \ref{nontrivial}.\\
i) The degeneracy of the eigenvalue $1$ of ${\cal U}_{0,0}(0)$ is
responsible for the existence of a non-trivial invariant sub-algebra of
observables which is the commutant of $h_0$.
\\

As in Section 2, we  generalize our result to the regime
$\lambda^2\tau\ra 0$, $\tau\ra 0$, by switching to the macroscopic time scale 
$T=t/(\lambda^2\tau)\ra \infty$. We first compute
\bea
&&\Gamma_\beta(B)=\lim_{\tau \ra 0} \frac{{\cal  U}_{0,0}(0)^{-1}\ T_\beta}{{\cal Z}(\beta)\tau^2}(B)
=-\frac{1}{2{\cal Z}(\beta)}({W^2}_{0,0}B+B{W^2}_{0,0})+\\
&&\frac{1}{{\cal Z}(\beta)}\sum_{m=1}^n\left\{ e^{-\delta_m\beta}\left(   
W_{m,0}BW_{0,m}-\frac{1}{2}({W^2}_{m,m}B+B{W^2}_{m,m})\right)+W_{0,m}BW_{m,0}\right\},
\nonumber
\eea
which, using the following formulas for $m\geq 1$
\be
W_{0,m}=V_m^*, \ \ W_{m,0}=V_m, \ \ W^2_{m,m}=V_mV_m^*, \ \ W^2_{0,0}=
\sum_{j=1}^nV_j^*V_j,
\ee  
to express the operators $W_{mm'}$ in terms of $V_m$, eventually becomes
\bea\label{4.59}
\Gamma_\beta(B)&=&\frac{1}{{\cal Z}(\beta)}\sum_{m=1}^ne^{-\beta\delta_m}
\left(V_{m}BV_{m}^*-
\frac{1}{2}(V_{m}V_{m}^*B+BV_{m}V_{m}^*)\right)\nonumber\\
& &\qquad\qquad\qquad\qquad + V_{m}^*BV_{m}-\frac{1}{2}(V_{m}^*V_{m}B+BV_{m}^*V_{m}).
\eea
We note here that this operator has the form of the dissipative part of a Lindblad generator. 
We'll come back to this operator $\Gamma_\beta$ in connection to the modelization in 
terms of Quantum Noises proposed in \cite{ap} and \cite{lm}, in the next Section.

\begin{cor}\label{tobewritten}
Assume the hypotheses of Theorem \ref{mtb}. Then
with $t/(\tau\lambda)^2=k\in\N$, 
\bea
&&\lim_{\tau\ra 0, \lambda^2\tau \ra 0 \atop 
t/(\tau\lambda)^2\in\N}{\cal U}_{0,0}(0)^{-t/(\tau\lambda)^2}
B_\beta( t/(\tau\lambda)^2,\lambda,\tau))=\\
&&\quad\quad \lim_{\tau\ra 0,  \lambda^2\tau \ra 0\atop 
t/(\tau\lambda)^2\in\N}
{\cal U}_{0,0}(0)^{-t/(\tau\lambda)^2}
{\cal U}_\beta(\lambda,\tau)^{t/\lambda^2}(B)=e^{t{\Gamma_\beta}^\#}(B),\nonumber
\eea
were $\Gamma_\beta(B)$ is defined in (\ref{4.59}).
\end{cor}
{\bf Proof:} We can simply repeat the arguments of the proof Theorem \ref{theo31}
once we note the following facts: 
i) The operator ${\cal U}_{0,0}(0)=e^{i\tau [h_0,\cdot ]}$ 
is unitary on $M_{d+1}(\C)$, with spectral projectors that are independent of $\tau$ 
as $\tau\ra 0$ and eigenvalues of the form $e^{i\tau \Delta_j}$. 
ii) Introducing $x=(\lambda\tau)^2$, (\ref{tbeta}) states that uniformly in $\tau$,
\be
{\cal U}_\beta(\lambda,\tau)={\cal U}_{0,0}(0)+xT_\beta(\tau)/(\tau^2{\cal Z(\beta)})
+O(x^2),
\ee
where $T_\beta(\tau)/\tau^2\ra \Gamma_\beta$ as $\tau\ra 0$. 
\ep

\subsection{Evolution of states}

Let us close this Section by briefly recalling some consequences
of these results about the evolution of states, i.e. trace one 
positive matrices. This is conveniently
done in our setup by using duality with respect to the scalar
product $\bra A | B\ket = \tr (A^* B)$.

If $\Gamma$ is the generator of the dynamics of observables, $B$ is
an observable and $\rho$ is a state, then for any $t\in\R$,
\be
\tr (\rho e^{t\Gamma}(B))=\tr (e^{t \Gamma_*}(\rho) B)
\ee
where the generator of the dynamics of the states is $\Gamma_*$
such that for all states $\rho$ and observables $B$,
\be
\tr ((\Gamma_*(\rho))^* B)=\bra \rho | \Gamma(B)\ket=
\bra \Gamma ^* \rho |B\ket.
\ee

In the particular case where
the observables $P_{jk}=|x_j\ket\bra x_k |$, with the notations of 
(\ref{deg}), form an orthonormal basis of eigenvectors of the restricted 
uncoupled evolution ${\cal U}_{0,0}$, the corresponding eigenprojectors 
are denoted by $\Pi_{jk}$ and act as
\be
\Pi_{jk}(B)=P_{jk}\tr (|x_k\ket\bra x_j| B)=P_{jk}\bra x_j |B x_k\ket_{\hil_0},
\ee
where the subscript $\hil_0$ denotes the scalar product within $\hil_0$.
Hence, the $\#$ operation on the operator $\Gamma$ with respect to the
projectors $\Pi_{jk}$ 
is given by
\be
\Gamma^\#(B)=\sum_{j,k}\Pi_{jk}\Gamma \Pi_{jk}(B)=
\sum_{j,k}\, |x_j\ket\bra x_k | \, \bra x_j |\Gamma (|x_j\ket\bra x_k |) 
x_k\ket_{\hil_0}\bra x_j | B x_k\ket_{\hil_0}.
\ee
Therefore, one computes that the corresponding generator of states, 
$(\Gamma^{\#})_*$ is given by
\be
\Gamma^{\#}_*(\rho)=\sum_{j,k}\, |x_j\ket\bra x_k | \, \bra x_k |
\Gamma (|x_j\ket\bra x_k |)^* 
x_j\ket_{\hil_0}\bra x_j | \rho x_k\ket_{\hil_0}.
\ee
Consequently,
\be
(\Gamma^\#)_*=\sum_{j,k}\Pi_{jk}\Gamma_* \Pi_{jk}=(\Gamma_*)^\#.
\ee

We note that states defined as functions of the Hamiltonian  $h_0$
of the small system form an invariant subspace of sets 
whose Markovian dynamics is characterized by the scalars
$\{ \bra x_j |\Gamma (|x_j\ket\bra x_j |) x_j\ket_{\hil_0} 
\}_{j=0,\cdots, d}$.

\section{Beyond the perturbative regime: $\lambda^2\tau=1$}
\setcounter{equation}{0}

We consider here the regime $\lambda^2\tau=1$, and $\tau\ra 0$ used in \cite{ap}
in their construction of the field of quantum noises.  It can be viewed as a regime where 
the weak limit scaling holds at the microscopic level, while, at the  
macroscopic level,  $T=t/(\tau\lambda^2)$ is kept finite.

As we saw in Corollaries \ref{3.5} and \ref{tobewritten} in the Schr\"odinger 
and Heisenberg pictures respectively, the small parameter that allows 
to make use of perturbation theory to compute the effective evolution is the
combination $\lambda^2\tau$. Therefore, we have to resort to a different technique
since our scaling imposes a non-perturbative regime. Our main tool will be
Chernoff's Theorem as we now explain.

\subsection{Schr\"odinger Evolution}

Let us start with the Schr\"odinger effective evolution under the following 
assumptions:  \\

{\bf H3:} Hypothesis {\bf H1} holds with $\ban$ a Hilbert space and 
$P$, $H(\lambda)=H(0)+\lambda W$ self-adjoint. \\

In the scaling adopted here, the number of interactions $n$ has to grow
like $n=t/\tau.$ This is in keeping with by the fact that in all cases 
considered so far, $n=t/(\lambda \tau)^2=t/\tau$. 
Note that the macroscopic time $T=\tau n = t$ is finite here.
Therefore, according to the analysis of Section 3, we are led to study
\be
PU(t/\tau,0)P = \left[Pe^{-i(\tau H(0)+\sqrt{\tau}W)}P \right]^{t/\tau},
\ \ \mbox{ as } \ \ \tau\ra 0, \ t/\tau\in\N^*.
\ee
This limit is easily computed by applying the following version
of Chernoff's Theorem, see e.g. \cite{br}, \cite{d} or \cite{p}, 
which suffices for our purpose:
\begin{thm}
Let $S(\tau)$ defined on a Banach space $\ban$ be such that
$S(0)=\un$, and $\|S(\tau)\|\leq 1$, for all $\tau\geq 0$.
If, 
$\lim_{\tau\ra 0} \tau^{-1}(S(\tau)-\un)=\Gamma$ in the strong sense
 exists in ${\cal L}(\ban)$
and generates a contraction semi-group, then
\be
s-\lim S(t/n)^n=e^{t\Gamma}.
\ee
\end{thm}
Now, it is easily checked that
\be
S(\tau):=Pe^{-i(\tau H(0)+\sqrt{\tau}W)}P \ \ \mbox{ on the subspace}\ \ P\ban
\ee
satisfies the first requirements. 
Then, by expanding the exponent and making use of the properties of $H(0)$ and $W$, 
we can write 
\be
S(\tau)=\left(\un-i\tau H(0)_P -\frac{\tau}{2}(W^2)_P+O(\tau^2)\right).
\ee
It thus implies
\be\label{spz}
S'(\tau)|_{\tau=0}=-iH(0)_P-\frac{(W^2)_P}{2}=\Gamma \in {\cal L}(P\ban).
\ee
Now $\Gamma$ is dissipative, since $\forall \ffi\in P\ban$
\be
\Re \bra \ffi |\Gamma \ffi\ket=-\Re \bra \ffi |PWQWP\ffi\ket/2=-\|QWP\ffi\|_\ban/2 \leq 0.
\ee
Hence, by Lumer-Phillips, see \cite{p}, $\Gamma$ generates a contraction
semigroup. Therefore
\begin{thm} Under the hypothesis {\bf H3}, for any $t>0$ fixed,
\be
s-\lim_{\tau\ra 0 \atop t/\tau\in \N}PU(t/\tau,0)P=s-\lim_{\tau\ra 0 \atop t/\tau\in \N}
 \left[Pe^{-i(\tau H(0)+\sqrt{\tau}W)}P \right]^{t/\tau}=e^{-t(iH(0)_P+\frac{(W^2)_P}{2})}.
\ee
\end{thm}
{\bf Remark:} Specializing to our model Hamiltonian, we get that the
effective dynamics on $P\ban$ is
\be
e^{-t(ih_0+\frac{1}{2}\sum_j V_j^*V_j)}.
\ee
Apart from the self-adjoint part $h_0$ stemming from the uncoupled evolution, the main 
difference with respect to the corresponding weak coupling 
result in Corollary \ref{3.5}, lies in the
absence of the $\#$ operation on the dissipative part $\frac{1}{2}\sum_j V_j^*V_j$ of
the generator. This prevents the spectral subspaces of $h_0$ from being invariant under
the effective dynamics.

\subsection{Heisenberg Evolution}

Let us now turn to the more interesting case of the Heisenberg dynamics of 
observables when the spins are at equilibrium at inverse temperature $\beta$. 
We assume the general hypothesis {\bf H0}, i.e. we stick to our matrix model, 
even though certain results below hold for more general situations.

\vspace{.3cm}

The analysis of Section 4 shows that the evolution of an observable 
$B\in M_{d+1}(\C)$ after $k$ repeated interactions reads
\be
B \mapsto B_\beta(k,\lambda,\tau)={\cal U}_\beta(\lambda,\tau)^k(B)
\ee
with ${\cal U}_\beta(\lambda,\tau)$ defined by (\ref{ubeta}), where we explicited the
dependence in $\tau$ in the notation. We want to apply Chernoff's Theorem again
to the operator valued function $\tau \mapsto {\cal U}_\beta(1/\sqrt{\tau},\tau)$
on ${\cal L}(M_{d+1}(\C))$. In order to check the first hypotheses we recall
the formula (see (\ref{4.9}))
\bea
 {\cal U}_\beta(\lambda,\tau)(B)&=&\tr_{\hil}\left((\un\otimes \omega_1(\beta))
U^{-1}(1,0)(B\otimes \un) U(1,0)\right)\\
&=&\sum_{q=0}^n\frac{e^{-\beta \delta_q}}{{\cal Z}(\beta)}\B(\tau)_{qq},\nonumber
\eea
where 
$\B(\tau)_{qq}=(U^{-1}(1,0)(B\otimes \un) U(1,0))_{qq}=P_q U^{-1}(1,0)(B\otimes \un) U(1,0) P_q$
according to the block notation (\ref{4.20}), with the corresponding orthogonal projectors $P_q$. 
Identifying $P_q\C^{(n+1)(d+1)}$ with $\hil_0=\C^{d+1}$, we deduce from the above formula that
${\cal U}_\beta(\lambda,\tau)$ is a contraction for any value of the parameters:
\bea
 \|{\cal U}_\beta(\lambda,\tau)(B)\|_{\hil_0}&\leq& 
\sum_{q=0}^n\frac{e^{-\beta \delta_q}}{{\cal Z}(\beta)}\|\B(\tau)_{qq}\|_{\hil_0}\\
&\leq& \sum_{q=0}^n\frac{e^{-\beta \delta_q}}{{\cal Z}(\beta)}\|P_qU^{-1}(1,0)(B\otimes \un) U(1,0)
 P_{q}\|_{\C^{(n+1)(d+1)}}\nonumber \\
&\leq & \sum_{q=0}^n\frac{e^{-\beta \delta_q}}{{\cal Z}(\beta)}\|(B\otimes \un)
\|_{\C^{(n+1)(d+1)}}=\|B\|_{\hil_0}.\nonumber
\eea
Moreover, ${\cal U}_\beta(1/\sqrt{\tau},\tau)|_{\tau=0}=\un$, so we are left with the
computation of the derivative w.r.t. $\tau$ at the origin. This involves the control 
of the operator $U_\tau(\lambda)$ (\ref{lop}) as $\tau\ra 0$ and 
$\lambda=1/\sqrt{\tau}\ra \infty$, as in the previous paragraph. Let us get estimates in a more
systematic way than above. So far, all
our estimates are derived for both $\lambda$ and $\tau$ going to zero or at most finite. 
However, the expansion of $U_\tau(\lambda)$ in powers of $\lambda$ is convergent, with
$\tau$ dependent coefficients we control sufficiently well. Indeed,  (\ref{cv})
yields
\be
U_\tau(\lambda)=e^{-i\tau H(0)}\Theta(\lambda, \tau)=\sum_{n\geq 0}
e^{-i\tau H(0)}\Theta_n(\lambda, \tau),
\ee
where $\Theta_n$ contains $n$ operators $W$ and satisfies
\be
\|\Theta_n(\lambda, \tau)\|=O((\tau\lambda)^n/n!).
\ee
Using the fact that $(\lambda \tau)^n=\tau^{n/2}\ra 0$ and that $W$ is 
off-diagonal with respect to $P$ and $Q$, we get that the replacement 
of $\lambda$ by $1/\sqrt{\tau}$ doesn't spoil the estimates as $\tau\ra 0$ given
in Proposition \ref{bkuk} and  Lemma  \ref{4.5}. Those together with the computation (\ref{4.59}) yield
\bea
{\cal U}_\beta(1/\sqrt{\tau},\tau)(B)&=&e^{i\tau h_0}Be^{-i\tau h_0}+({\cal Z(\beta)}\tau)^{-1}
T_\beta(\tau)(B)+O(\tau^2)\nonumber\\
&\equiv& e^{i\tau h_0}Be^{-i\tau h_0}+\tau \Gamma_\beta(B) +O(\tau^2),
\eea
where, see (\ref{4.59}),
\bea
\Gamma_\beta(B)&=&\frac{1}{{\cal Z}(\beta)}\sum_{m=1}^ne^{-\beta\delta_m}
\left(V_{m}BV_{m}^*-
\frac{1}{2}(V_{m}V_{m}^*B+BV_{m}V_{m}^*)\right)\nonumber\\
& &\qquad\qquad\qquad\qquad + V_{m}^*BV_{m}-\frac{1}{2}(V_{m}^*V_{m}B+BV_{m}^*V_{m}).
\eea
Hence, the derivative at the origin exists and is given by
\be
{\cal U}_\beta(1/\sqrt{\tau},\tau)'(B)|_{\tau=0}=i[h_0,B]+\Gamma_\beta(B).
\ee

We recognize at once that $\Gamma_\beta(B)$ is the dissipative part of a Lindblad operator 
of the form
\be\label{lind}
\sum_{j=1}^{2m}L_jBL_j^*-\frac{1}{2}\left(L_{j}L_{j}^*B+BL_{j}L_{j}^*\right)
\ee
with
\be
L_j=\frac{e^{-\beta\delta_j/2}}{\sqrt{{\cal Z}(\beta)}}V_j, \ 1\leq j\leq m \ \ \
\mbox{ and } \ \ \ L_j=\frac{1}{\sqrt{{\cal Z}(\beta)}}V_j^*, \ m+1\leq j\leq 2m.
\ee
By the Theorem of Lindblad, see e.g. \cite{af}, we know that  
\be
i[h_0,B]+\Gamma_\beta(B)
\ee
generates a completely positive semigroup of contractions.
Therefore, we are in a position to apply Chernoff's theorem to eventually get
\begin{thm}\label{thlin} Assume hypothesis H0 where $\hil_0$ is a separable
Hilbert space and $h_0$, the $V_j$'s and $B$ are bounded on $\hil_0$. Let 
$B_\beta(t/\tau,1/\sqrt\tau,\tau)$ be defined by (\ref{bkt}), 
${\cal U}_\beta(\lambda,\tau)$ is defined by proposition \ref{bkuk} and the 
Remark following it. Then
\be
s-\lim_{\tau\ra 0 \atop t/\tau\in\N} B_\beta(t/\tau,1/\sqrt\tau,\tau)=
s-\lim_{\tau\ra 0 \atop t/\tau\in\N} {\cal U}_\beta(1/\sqrt{\tau},\tau)^{t/\tau}(B)=
e^{t(i[h_0,\cdot]
+\Gamma_\beta(\cdot))}(B)
\ee
with a Lindblad generator $i[h_0,\cdot]+\Gamma_\beta(\cdot)$ explicited in (\ref{lind})
\end{thm}
{\bf Remarks:} \\
i) Let us make a comparison of the above with the results of \cite{ap}, 
Section IV.2, which concern similar generators as ours. More precisely, (\ref{hamil}) 
corresponds to a particular case of the Hamiltonian of eq. (15) in \cite{ap}, with
$D_{ij}=0$, $\forall i,j$. 
In \cite{ap}, the choice of time scale $\tau$ and coupling 
$\lambda$ is such that $\lambda^2\tau=1$, $\tau\ra 0$.
A supplementary structure is present in that work which
consists in making the suitably renormalized spins 
forming the chain merge in the limit
$\tau\ra 0$ to yield a heat bath represented by a Fock space
of quantum noises. The limit $\tau\ra 0$ performed in the language 
adopted in \cite{ap} exists and yields a quantum Langevin 
equation for the whole limiting system consisting in the original 
small system in interaction with a field of quantum noises. 
When restricted to $\hil_0$, the effective dynamics of observables at zero temperature
corresponds to a contraction semigroup generated by  
\be
\Gamma_{\infty}(\cdot )=i[h_0,\cdot]+\sum_{m=1}^n\left(V_{m}^* \cdot V_{m}-
\frac{1}{2}(V_{m}^*V_{m} \cdot + \cdot V_{m}^*V_{m})\right),
\ee
which coincides with Theorem \ref{thlin} at $\beta=\infty$.\\
ii) The generator $\Gamma^{\beta}$ coincides with the generator 
(\ref{4.59}) obtained in Corollary \ref{tobewritten} in the scaling 
$\lambda^2\tau\ra 0$, $\tau\ra 0$, modulo the $\#$ operation, 
which appears as a trade mark of the perturbative regime.

\subsection{The Continuous Limit}

For completeness, we mention here the easier cases of continuous limit characterized by 
$\tau\ra 0$ and $\lambda$ constant. The omitted proof are quite analogous to those of 
the previous Section. \\

First considering the Schr\"odinger picture, we get
\begin{prop} Assume the hypothesis {\bf H3} holds and fix $\lambda=1$. 
Then, 
\be
s-\lim_{\tau\ra 0 \atop t/\tau\in \N}PU(t/\tau,0)P=s-\lim_{\tau\ra 0 \atop t/\tau\in \N}
 \left[Pe^{-i\tau( H(0)+W)}P \right]^{t/\tau}=e^{-itH(0)_P}.
\ee
\end{prop}
The Heisenberg evolution also yields a unitary effective evolution in the continuous
limit:
\begin{prop} Consider the matrix model of Section 2 and fix $\lambda=1$. 
Then, 
\be
\lim_{\tau\ra 0 \atop t/\tau\in \N}B_\beta(t/\tau, \tau,1)=e^{i[h_0,\cdot]}(B).
\ee
\end{prop}

\vspace{.5cm}

In order to make explicit the results of Section \ref{sechei}, we provide below a 
detailed analysis of the case $d=n=1$.

\setcounter{equation}{0}
\section{The case $d=n=1$}
In that Section, we focus on the first non-trivial case where the
small system lives on $\C^2$ and the heat bath is formed by a chain 
of spins $1/2$. We provide explicit formulas for $T_\beta$ and $T_\beta^\#$
which are valid for any coupling operator $V$ appearing in (\ref{defint}).
We further diagonalize the restriction of $T_\beta$ to the degenerate 
subspace $Ker({\cal U}_{0,0}(0)-1)$ in order to determine the 
subalgebra of observables invariant under the effective dynamics in the
weak coupling limit (keeping $\tau$ fixed).
\\

For $\hil_0=\C^2$, $\hil=\otimes_{j\geq 1}\C^2$, we write for 
$t\in [\tau k-1, \tau k[$ in $\hil_0\otimes C^2_k$,  
\be
H(t, \lambda)= H(\lambda)=H(0)+\lambda W,
\ee
where 
\be
H(0)=h_0\otimes\un+\un\otimes\delta a^* a, \ \ \ W=V^*\otimes a
  +V\otimes a^*.
\ee
We choose, without loss of generality, $h_0=\eps\sigma_z, \eps\neq 0$,
so that we have in the 
ordered basis $\{\omega \otimes \omega, x\otimes\omega,
  \omega \otimes x, x \otimes x\}$ 
\be\label{63}
H(\lambda)=\pmatrix{\eps\sigma_z & \lambda V^* \cr \lambda V &
  \delta \un +\eps\sigma_z}, \ \ \ \mbox{with the convention} \ \ \
\sigma_z=\pmatrix{-1 & 0 \cr 0& 1}. 
\ee

Specifying the results of the previous sections to the case under study, we can write,
uniformly in $\beta$, as $\lambda\ra 0$, 
\be
{\cal U}_{\beta}(\lambda)={\cal U}_{0,0}(0)+\lambda^2 \frac{T_\beta}{1+e^{-\delta\beta}} 
+O(\lambda^4),
\ee
with
\bea
T_\beta(B)&=&F_{0,1}(-\tau)BF_{1,0}(\tau)+G_{0,0}(-\tau)Be^{-i\tau H_{0,0}(0)}+
e^{i\tau H_{0,0}(0)}BG_{0,0}(\tau)\\
&+&e^{-\delta\beta}(F_{1,0}(-\tau)BF_{0,1}(\tau)
+G_{1,1}(-\tau)Be^{-i\tau H_{1,1}(0)}+
e^{i\tau H_{1,1}(0)}BG_{1,1}(\tau)).\nonumber
\eea

We use the norm induced by the scalar product $\bra A, B\ket =\tr (A^*B)$, i.e.
the Hilbert-Schmidt norm.
As easily verified, an orthonormal basis of eigenvectors for the unitary operator 
${\cal U}_{0,0}(0)(\cdot)=e^{i\tau\eps\sigma_z}\,\cdot \, e^{-i\tau\eps\sigma_z}$,
with associated eigenvalues, is provided by 
\be
\{\hat{\un}, \hat\sigma_z, \sigma_-,\sigma_+\}\longleftrightarrow 
\{1,1,e^{-2i\tau\eps}, e^{2i\tau\eps}\},
\ee
where
\be
\sigma_+=\pmatrix{0 & 0 \cr 1& 0}, \ \ \sigma_z=\pmatrix{0 & 1 \cr 0& 0},
\  \mbox{ 
$\hat\un=\un/\sqrt 2$ \, and \,  $\hat\sigma_z=\sigma_z/\sqrt 2$}. 
\ee
Let us compute $T_\beta$ restricted to the subspace 
$\mbox{Ker}\,({\cal U}_{0,0}(0)-1)$ appearing in $T_\beta^\#$.
\begin{lem}\label{7.1}
With respect to the orthonormal basis $\{\hat{\un}, \hat\sigma_z\}$,
and with the notation 
\be
A^{OD}=\pmatrix{0&b\cr c&0}\,\,\, \mbox{ if }\,\,\, 
A=\pmatrix{a&b\cr c&d}\in M_2(\C)
\ee
we have
\be
T_\beta|_{\{\hat{\un}, \hat\sigma_z\}}=
\pmatrix{0 &   {T_\beta}_{1,2}\cr 0 & {T_\beta}_{2,2}},
\ee
where
\bea\label{valb}
{T_\beta}_{1,2}&=&\left(|(F_{1,0}^{OD})_{2,1}|^2-|(F_{1,0}^{OD})_{1,2}|^2\right)
(1-e^{-\delta\beta})\nonumber \\
{T_\beta}_{2,2}&=&-(\|F_{1,0}^{OD}\|^2+e^{-\delta\beta}
\|F_{0 ,1}^{OD}\|^2)=-\|F_{1,0}^{OD}\|^2
(1+e^{-\delta\beta})\leq 0.
\eea
Furthermore, if in (\ref{defint}) $V=\pmatrix{\bf a &\bf b\cr\bf c &\bf d}$,
\be\label{expcal}
F_{1,0}(\tau)=-i\pmatrix{e^{i\tau(\eps-\delta)}\int_0^\tau e^{is\delta}ds 
\,{\bf a} & 
e^{i\tau(\eps-\delta)}\int_0^\tau e^{is(\delta-2\eps)}ds \, {\bf b}\cr
e^{-i\tau(\eps+\delta)}\int_0^\tau e^{is(\delta+2\eps)}ds \, {\bf c}& 
e^{-i\tau(\eps+\delta)}\int_0^\tau e^{is\delta}ds \, {\bf d}}.
\ee
\end{lem}
{\bf Proof:} The first column is proportional to $T_\beta(\un)=0$. 
The second column of the matrix is given by 
\be\label{vec}
\frac{1}{2}\pmatrix{\tr (T_\beta(\sigma_z))\cr\tr (\sigma_zT_\beta(\sigma_z)) },
\ee
where, dropping the positive argument $\tau$ in $F$ and further making use of  
(\ref{ftau}) and (\ref{gtau}),
\bea
T_\beta(\sigma_z))&=&F_{1,0}^*\sigma_zF_{1,0}+G_{0,0}^*\sigma_ze^{-i\tau H_{0,0}(0)}+
e^{i\tau H_{0,0}(0)}\sigma_zG_{0,0}\nonumber\\
&+& e^{-\delta\beta}\left(F_{0,1}^*\sigma_zF_{0,1}+G_{1,1}^*\sigma_ze^{-i\tau H_{1,1}(0)}+
e^{i\tau H_{1,1}(0)}\sigma_zG_{1,1}\right).
\eea
Further making use of the cyclicity of the trace, $[\sigma_z,H_{n,n}(0)]=0$, 
$\sigma_z^2=\un$  and of (\ref{ftau}) and (\ref{gtau}) again, we can write
\bea
\tr(\sigma_zT_\beta(\sigma_z)))&=&\tr(\sigma_zF_{1,0}^*\sigma_zF_{1,0})-
\tr(F_{1,0}^*F_{1,0})\nonumber\\
&+&e^{-\delta\beta}(\tr(\sigma_zF_{0,1}^*\sigma_zF_{0,1})-
\tr(F_{0,1}^*F_{0,1})).
\eea
Explicit computations on $2\times 2$ matrices yields the first equality in 
(\ref{valb}). 
Let us  turn to  (\ref{expcal}).
From the definitions (\ref{deff}) and (\ref{defint}), we have
\be
F(\tau)=\pmatrix{\0 &F_{0,1}(\tau) \cr F_{1,0}(\tau) & \0},
\ee
where 
\bea
F_{0,1}(\tau)&=&-i\int_0^\tau e^{-i(\tau-s)H_{0,0}(0)}V^*e^{-isH_{1,1}(0)}ds,\\
F_{1,0}(\tau)&=&-i\int_0^\tau e^{-i(\tau-s)H_{1,1}(0)}Ve^{-isH_{0,0}(0)}ds.
\eea
By explicit computations with $V$ as in the statement, we obtain
\bea
F_{0,1}(\tau)&=&-i\pmatrix{e^{i\tau\eps}\int_0^\tau e^{-is\delta}ds \,
\overline{\bf a} & 
e^{i\tau\eps}\int_0^\tau e^{-is(\delta+2\eps)}ds \,\overline{\bf c}\cr
e^{-i\tau\eps}\int_0^\tau e^{-is(\delta-2\eps)}ds \,\overline{\bf b}& 
e^{-i\tau\eps}\int_0^\tau e^{-is\delta}ds \,\overline{\bf d}}\\
F_{1,0}(\tau)&=&-i\pmatrix{e^{i\tau(\eps-\delta)}\int_0^\tau e^{is\delta}ds 
\,{\bf a} & 
e^{i\tau(\eps-\delta)}\int_0^\tau e^{is(\delta-2\eps)}ds \, {\bf b}\cr
e^{-i\tau(\eps+\delta)}\int_0^\tau e^{is(\delta+2\eps)}ds \, {\bf c}& 
e^{-i\tau(\eps+\delta)}\int_0^\tau e^{is\delta}ds \, {\bf d}},
\eea
which yields the expression for ${T_\beta}_{i,j}$.\\
By similar manipulations we get
\bea
\tr(\un T_\beta(\sigma_z))&=&\tr(F_{1,0}^*\sigma_zF_{1,0})-
\tr(\sigma_zF_{1,0}^*F_{1,0})\nonumber\\
&+&e^{-\delta\beta}(\tr(F_{0,1}^*\sigma_zF_{0,1})-
\tr(\sigma_zF_{0,1}^*F_{0,1})).
\eea
Now, for any $F\in M_2(\C)$,
\be
\tr(\sigma_z(FF^*-F^*F))=2(|F_{21}|^2-|F_{12}|^2),
\ee
so that we get the first line of (\ref{valb}).
\hfill \ep

\vspace{.3cm}

We also need to compute $\tr (\sigma_-T_\beta(\sigma_+))$ and 
$\tr (\sigma_+T_\beta(\sigma_-))$ to get $T_\beta^\#$.
\begin{lem} By explicit computation and Lemma \ref{symt}, we have
\bea
\tr (\sigma_-T_\beta(\sigma_+))&=&\overline{\tr
  (\sigma_+T_\beta(\sigma_-))}\nonumber\\
&=& (F_{1,0})_{1,1}\overline{(F_{1,0})_{2,2}}+e^{i\tau\eps}(
(G_{0,0})_{1,1}+\overline{(G_{0,0})_{2,2}})\\
&+&e^{-\delta\beta}\left((F_{0,1})_{1,1}\overline{(F_{0,1})_{2,2}}+
e^{i\tau\eps}(
e^{i\tau\delta}(G_{1,1})_{1,1}+e^{-i\tau\delta}
\overline{(G_{1,1})_{2,2}})\right).\nonumber
\eea
\end{lem}

It remains to diagonalize the restriction of $T_\beta$ to span($\hat \un, \hat \sigma_z$)
to have a complete description of the generator of the effective evolution. 
Introducing
\be
\mu={T_\beta}_{1,2}, \ \ \ 
\nu={T_\beta}_{2,2},
\ee we actually get by perturbation theory, 
\begin{lem}\label{perpr}
Assume $\eps\tau\notin \Z\pi$. Then, for $\lambda>0$, 
there exists a continuous set of 
eigenprojectors and eigenvalues of ${\cal U}_\beta(\lambda,\tau)$ 
denoted respectively by $\{\Pi_j(\lambda)\}_{j=1,\cdots, 4}$ and
$\{u_j(\lambda)\}_{j=1,\cdots, 4}$ such that
\bea
&&u_1(\lambda)=1+O(\lambda^4),\nonumber\\&& u_2(\lambda)=1-
\lambda^2\|F_{1,0}^{OD}\|^2+O(\lambda^4),\nonumber\\
&&u_3(\lambda)=e^{2i\tau\eps}+\lambda^2\tr(\sigma_-T_\beta
(\sigma_+))+O(\lambda^4),\nonumber\\ 
&&u_4(\lambda)=e^{-2i\tau\eps}+\lambda^2\tr(\sigma_+T_\beta
(\sigma_-))+O(\lambda^4),
\eea
and 
\bea
&&\Pi_1(\lambda)(B)=\frac{\tr((\un-\frac{\mu}{\nu} \sigma_z)B)}{2}\un +O(\lambda^2), 
 \,\,\,\,
\Pi_2(\lambda)(B)= \frac{\tr(\sigma_z B)}{2}(\frac{\mu}{\nu} \un+\sigma_z)
+O(\lambda^2)\nonumber \\
&&\Pi_3(\lambda)(B)= \tr(\sigma_- B)\sigma_+
+O(\lambda^2), \,\, \ \ \Pi_4(\lambda)(B)= \tr(\sigma_+ B)\sigma_-
+O(\lambda^2).
\eea
Moreover, $\Pi_0:=\Pi_1(0)+\Pi_2(0)$, $\Pi_3(0)$ and $\Pi_4(0)$ are the
spectral projectors of ${\cal U}_{0,0}(0)$ and $\{\Pi_j(0)\}_{j=1,\cdots,4}$ are those
of $T_\beta$.
\end{lem}

Hence, we obtain the 
\begin{prop}\label{nontrivial} Let $t/\lambda^2=k\in\N$, and consider the Hamiltonian (\ref{63}). Then
\be
\lim_{\lambda\ra 0}{\cal U}_{0,0}(0)^{-t/\lambda^2}B( t/\lambda^2,\beta,\lambda)=
\lim_{\lambda\ra 0}{\cal U}_{0,0}(0)^{-t/\lambda^2}
{\cal U}_\beta(\lambda, \tau)^{t/\lambda^2}(B)=e^{t\Gamma^w_\beta}(B),
\ee
were
\bea
\Gamma_\beta^w&=&\frac{1}{1+e^{-\delta\beta}}
(-\|F_{1,0}^{OD}\|^2\Pi_2(0)\\
&+&e^{-2i\tau\eps}\tr(\sigma_-T_\beta
(\sigma_+))\Pi_3(0) +e^{2i\tau\eps}\tr(\sigma_+T_\beta
(\sigma_-))\Pi_4(0)).\nonumber
\eea
\end{prop}
The dynamics of any observable is thus fully determined from these formulas.

\vspace{ .3cm}

\noindent
{\bf \Large Acknowledgements:}

\vspace{ .3cm}
We wish to thank Laurent Bruneau for a careful and critical reading of
the manuscript and Claude-Alain Pillet for useful discussions.


\begin{thebibliography}{xxxxxxx}
 \bibitem[AF]{af} Alicki, R., Fannes, M: Quantum Dynamical Systems, 
Oxford University Press, 2001.
\bibitem[AP]{ap} Attal, S., Pautrat, Y.: `` From repeated to
  continuous quantum interactions``, Preprint (2003). 
\bibitem[BR]{br} Brattelli O., Robinson D. ``Operator Algebras and Quantum 
Statistical Mechanics II'', Texts and Monographs in Physics, Springer,
New York, Heidelberg, Berlin, 1981.
\bibitem[D1]{d0} Davies, E.B.: ``Markovian master equations,'' {\it Comm. Math. Phys.}, 
{\bf 39}, 91--110, (1974).
\bibitem[D2]{d} Davies, E.B.: One-Parameter Semigroups, Academic Press, 1980.
\bibitem[DJ]{dj} Derezinski, J., Jaksic, V.: `` On the Nature of Fermi Golden Rule 
of Open Quantum Systems``, {\it J.Stat.Phys.}  {\bf 116}, (2004), 411-423.
\bibitem[DS]{ds} Davies, E.B., Spohn, H.: ``Open Quantum Systems with Time-Dependent 
Hamiltonians and Their Linear Response'', {\it J.Stat.Phys.} {\bf 19}, 511-523, (1978). 
 \bibitem[K]{k} Kato, T.: Perturbation Theory for Linear Operators, Springer, (1980).
\bibitem[LS]{ls} Lebowitz, J. and Spohn, H., ``Irreversible Thermodynamics for 
Quantum Systems Weakly Coupled to Thermal Reservoirs'', {\it Adv.Chem.Phys.} {\bf 39}, 
109-142, (1978). 
\bibitem[LM]{lm} J.M. Lindsay and H. Maassen, ``Stochastic Calculus for Quantum Brownian 
  Motion of a non-minimal variance'' In: Mark Kac Seminar of probability in Physics, 
Syllabus 1987-1992, CWI Syllabus 32, Amsterdam, (1992).  
 \bibitem[Paz]{p} Pazy, A: Semigroups of Linear Operators and Applications to Partial 
Differential Equations, Springer, 1983.
 \end{thebibliography}
\end{document}